\documentclass{jfm}

\usepackage{graphicx}
\usepackage{newtxtext}
\usepackage{newtxmath}
\usepackage{natbib}
\usepackage{subfigure}
\usepackage{hyperref}
\usepackage{booktabs}
\usepackage{multirow}
\usepackage{threeparttable}
\usepackage{float}

\hypersetup{
    colorlinks = true,
    urlcolor   = blue,
    citecolor  = black,
}

\newcommand{\RomanNumeralCaps}[1]
\linenumbers

\title{A compressible Reynolds-averaged mixing model considering turbulent entropy and heat flux}

\author{Hansong Xie\aff{1},
  Tengfei Luo\aff{4,5}, Yaomin Zhao\aff{1}\corresp{\email{yaomin.zhao@pku.edu.cn}}, Yousheng Zhang\aff{1,2,3}\corresp{\email{zhang\_yousheng@iapcm.ac.cn}}
 \and Jianchun Wang\aff{4}}

\affiliation{\aff{1}HEDPS, Center for Applied Physics and Technology, and College of Engineering, Peking University, Beijing 100871, China
\aff{2}Institute of Applied Physics and Computational Mathematics, Beijing 100094, China
\aff{3}National Key Laboratory of Computational Physics, Beijing 100088, China
\aff{4}Department of Mechanics and Aerospace Engineering, Southern University of Science and Technology, Shenzhen 518055, China
\aff{5} Center for Ocean Research in Hong Kong and Macau, and Division of Environment and Sustainability, Hong Kong University of Science and Technology, Hong Kong, China}

\begin{document}
\maketitle

\begin{abstract}
In typical nature and engineering scenarios, such as  supernova explosion and  inertial confinement fusion,  mixing flows induced by hydrodynamics interfacial instabilities are essentially compressible.
Despite their significance, accurate predictive tools for these compressible flows remain scarce.
For engineering applications, the Reynolds-averaged Navier-Stokes (RANS) simulation stands out as the most practical approach due to its outstanding computational efficiency.
However, the majority of RANS mixing studies reported have concentrated on incompressible scenarios, with quite limited attention given to compressible cases.
Moreover, most of the existing RANS mixing models demonstrate significantly inaccurate predictions for compressible mixing flow. 
This study develops a novel compressible RANS mixing model by incorporating  physical compressibility corrections into the incompressible $K-L-\gamma$  mixing transition model recently proposed by  Xie \emph{et al.} (\emph{J. Fluid Mech.}, 1002, A31, 2025).
Specifically, taking the density-stratified  Rayleigh-Taylor mixing flows as representative compressible cases, we firstly analyze the limitations of the existing model for compressible flows, based on  high-fidelity data and  local instability criteria.
Subsequently, the equation of state for a perfect gas and the thermodynamic Gibbs relation are employed  to derive  comprehensive compressibility corrections.
The crucial turbulent entropy and heat flux are integrated into the closure of the key turbulent mass flux term of the turbulent kinetic energy equation.
These corrections enable the model to accurately depict compressible mixing flows.
Systematic validations confirm the efficacy of the proposed modeling scheme.
This study offers a promising strategy for modeling compressible mixing flows, paving the way for more accurate predictions in complex scenarios.
\end{abstract}


\section{Introduction} \label{Introduction}
This investigation focuses on  mixing flows induced by hydrodynamic interfacial instabilities, such as the Rayleigh-Taylor (RT) \citep{rayleigh1882investigation,taylor1950instability}
and  Richtmyer-Meshkov (RM) \citep{richtmyer1960taylor,meshkov1969instability} instabilities.
The RT instability arises when a denser fluid is accelerated by a lighter one under  constant acceleration, while the RM instability emerges if the acceleration is pulsing regardless of the shocking direction.
In both cases the misalignment between density and pressure gradients drives interfacial perturbation growth.
Baroclinic vorticity  plays a crucial role in amplifying interfacial perturbations while simultaneously entraining surrounding irrotational fluids into the mixing zone.
Mixing is frequently intensified by  the Kelvin-Helmholtz instability \citep{kelvin1871hydrokinetic,helmholtz1868on}, and subsequently transitions to turbulence.
This interfacial mixing phenomenon is observed in systems spanning multiple scales, from cosmic-scale supernova explosions to laboratory-scale applications in scramjet propulsion and inertial confinement fusion (ICF) \citep{zhou2017Rayleigh}.
The prevalence of these mixing processes across such diverse physical systems underscores their fundamental importance in both natural phenomena and engineering applications.

\begin{figure} 
\centering
\subfigure{
\includegraphics[width=0.35\textwidth]{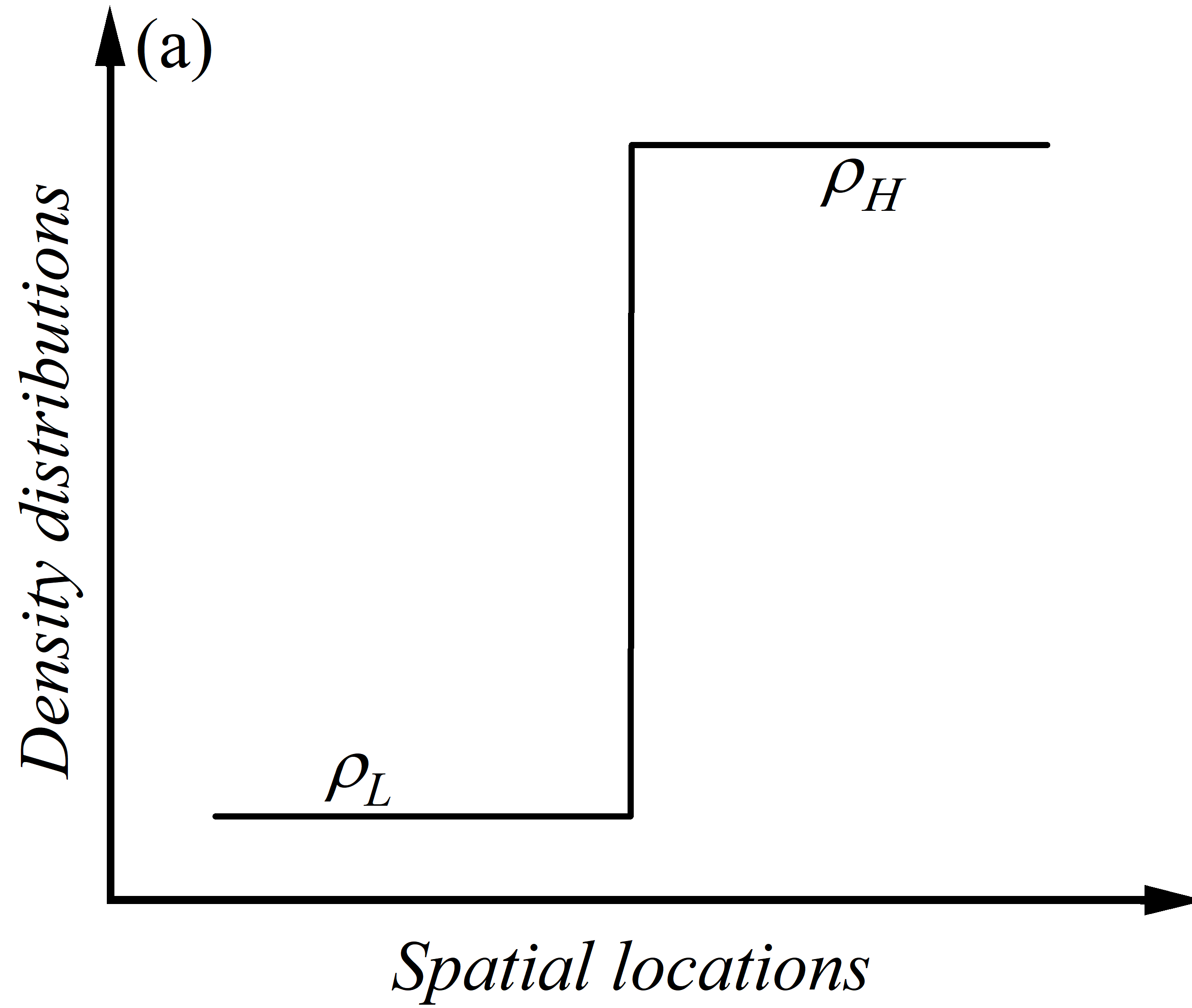}}\hspace{10mm}
\subfigure{
\includegraphics[width=0.35\textwidth]{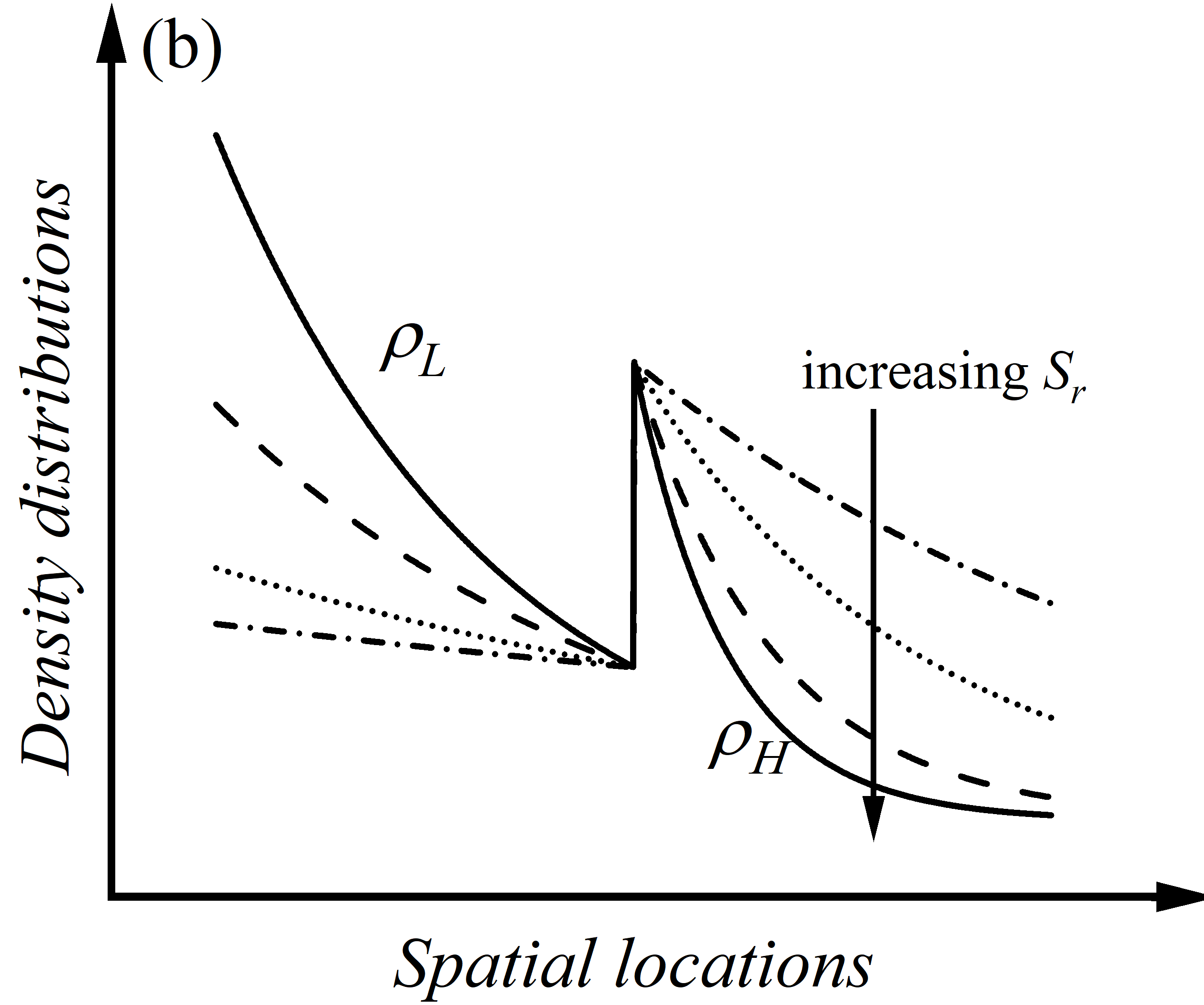}} 
\caption{Schematic diagrams of the initial density distributions for (a) the incompressible  and (b) the density-stratified compressible RT flows. } 
\label{sketch map}
\end{figure}

The aforementioned  mixing flows essentially demonstrate significant compressibility effects across numerous scenarios.
Taking ICF as an example, the internal flow can be characterized as a highly nonlinear compressible system progressing through four characteristic stages: shock-transit, acceleration, deceleration, and peak compression \citep{Craxton2015Direct}. 
The first stage typically initiates with a sequence of  laser pulses of increasing intensities irradiating the ICF target.
These pulses generate pressure waves in the plasma, launching a sequence of shock waves that propagate into and compress the target.
During this stage, surface perturbations on the shells  evolve as a result of a RM-like instability, due to imperfections in fabrication and laser-beam nonuniformity.
After the shock reaches the inner surface, the acceleration stage begins, characterized by an outward-moving rarefaction wave and inward acceleration of the target shells.
The subsequent deceleration stage sees continued shell compression with a gradually decreasing speed through centripetal inertia, with kinetic energy converted into internal energy, resulting in increase of pressure and temperature.
The compression process culminates in  peak compression phase, where the fuel reaches the maximum compression level, yielding central hot spots with extraordinarily high temperature and density  due to compressibility.
Throughout these stages, hydrodynamic instabilities, particularly the  RT and RM instabilities, amplify surface perturbations, compromising shell integrity and compression efficiency. 
The resulting mixing flows dissipate substantial energy, cooling the fuel and potentially causing ignition failure.
Notably, the density profiles within each shell vary radially as the radial location $r$, as illustrated in figure \ref{sketch map}(b), contrasting with the constant density profiles of incompressible cases in figure \ref{sketch map}(a). 
This ignition process underscores the critical importance of studying compressible mixing flows.

Current studies on compressible mixing flows primarily focus on those induced by  RT instability, where compressibility effects stem from two distinct origins \citep{Gauthier2010Compressibility}.
The first is the static compressibility, which results from the fluid’s variable density.
In an acceleration field, it causes density stratification and gradient for light and heavy fluids.
The second is dynamic compressibility, an effect fundamentally associated with the finite speed of sound.
For a perfect gas, it can be characterized by the adiabatic indices $\Gamma$ in the equation of state (EOS). 
The dynamic compressibility mainly depends  on the inherent properties of the fluid, while existing numerical and theoretical investigations have demonstrated that compressibility in RT flows is predominantly governed by static compressibility \citep{Livescu2004Compressibility,Jin2005Rayleigh,Xue2010Destabilizing,Gauthier2013Compressibility,Gauthier2017Compressible}. 
Therefore, the present study concentrates on static compressibility, specifically density-stratified compressible RT mixing flows.

The classical incompressible RT mixing, as illustrated in figure \ref{sketch map}(a), maintains  initial constant density distributions within both light and heavy fluid layers.
In contrast, the density-stratified compressible RT mixing initially displays significant density gradient, as depicted in figure \ref{sketch map}(b).
Under the isothermal condition, this density distribution exhibits exponential variation with the spatial locations. 
The stratification strength is commonly quantified using the density stratification parameter \citep{Luo2022Mixing}:
\begin{equation} \label{Sr}
    S_r\equiv\frac{ g L_r}{RT_r/M_r},
\end{equation}
with  $g$, $L_r$, $T_r$, $M_r$, and $R$ representing the gravitational acceleration, characteristic length, temperature, mole mass, and universal gas constant, respectively. 
The  larger  $S_r$ corresponds to a steeper density gradient, indicating a stronger compressibility effect. 
This fundamental difference highlights the distinctive behavior of compressible versus incompressible RT mixing.

Existing studies have revealed prominent differences between density-stratified compressible RT flows and their incompressible counterparts.
Owing to the  initial density stratification, the growth rate of RT mixing is substantially affected \citep{Xue2010Destabilizing}.
\cite{George2005Self} have emphasized that self-similar rates differ markedly between compressible and incompressible scenarios, noting that the quadratic growth rate with time (denoted as $t$) observed in incompressible cases cannot be directly applied to compressible flows.
This perspective has been corroborated by high-fidelity (HiFi) numerical simulations conducted by \cite{Gauthier2017Compressible} and \cite{Luo2022Mixing}, which confirm that the $t^2$-scaling law does not hold for density-stratified compressible RT mixing flows. 
Furthermore, density stratification enhances asymmetric evolution of the mixing layers on the light and heavy fluid sides.
Additionally, these studies have revealed that spatial profiles of density and temperature are significantly influenced by initial density stratification, exhibiting complex distributions that contrast with the monotonic variations seen in incompressible cases.

Energy transport mechanisms also exhibit fundamental differences between density-stratified compressible and incompressible RT flows.
For the compressible cases, kinetic energy generation stems from two primary sources: potential energy conversion and pressure-dilatation work \citep{Zhao2020Kinetic,Luo2022Mixing}. 
When the  $S_r$ is small, the kinetic energy derives predominantly from potential energy, which is similar to the incompressible situation.
However, for larger $S_r$, pressure-dilatation work becomes the dominant mechanism for kinetic energy generation. 
Moreover, the kinetic energy is subsequently converted back into potential energy.
These observations demonstrate that the underlying energy transport physics differs substantially between compressible and incompressible RT flows \citep{Zhao2022Scale}. 

While considerable progress has been made in understanding the fundamental mechanisms of density-stratified compressible  RT flows, modeling efforts remain limited. 
Current researches primarily focus on two approaches: the buoyancy-drag model \citep{Jin2005Rayleigh,Fu2022Nonlinear} and the large-eddy simulation (LES) techniques \citep{Luo2024Fourier}.
However, significant limitations persist. 
The buoyancy-drag model, while computationally efficient, oversimplifies the physics and fails to capture essential flow characteristics required for engineering applications.
Conversely, LES provides more detailed flow information but at prohibitively high computational costs, making it impractical for routine engineering predictions.
This disparity underscores the urgent need to develop alternative modeling approaches that can provide acceptable computational accuracy at an affordable cost, which is a critical step toward practical engineering applications of compressible RT flow predictions.

Reynolds-averaged Navier-Stokes (RANS) simulation has been widely adopted in industrial applications owing to its favorable balance between computational efficiency and cost.
A number of RANS mixing models have been proposed  for interfacial flow predictions, such as the $K-L$ \citep{Dimonte2006K-L,Morgan2015Large,Kokkinakis2019Modeling,zhang2020Methodology}, $K-\epsilon$ \citep{MornLpez2013MulticomponentRN,Xie2021Predicting}, and BHR models \citep{Denissen2014tilted,Xie2021Unified}.
However, most of the existing RANS models predominantly address incompressible flows.
The fundamental differences between compressible and incompressible RT mixing flows, as previously discussed, render existing RANS models inadequate for compressible scenarios. 
Indeed, compressible RANS modeling remains largely unexplored --- a critical gap this study seeks to address.
Our approach builds upon the recently proposed  $K-L-\gamma$ mixing transition model \citep{Xie2025intermittency}, which  has been shown to have good performance for capturing the mixing transition process in incompressible flows. 
Through systematic compressibility modifications, we extend this model to accurately capture both the transition dynamics \citep{Qi2024vortex} and compressibility effects in compressible mixing flows.

This paper is organized as follows. 
Section \ref{Sec:Governing equations and baseline model} presents the governing equations of the baseline $K-L-\gamma$ mixing transition model.
The performance of this baseline model for density-stratified compressible RT flows is then evaluated in section \ref{Convective instability in density-stratified compressible RT flow}. 
Through comparison with HiFi simulation data and application of local instability criteria, 
the fundamental limitations of the baseline model applied to compressible RT flows are systematically analyzed.
Section \ref{Compressibility corrections} details the theoretical derivation of the proposed compressibility modifications.
These developments are subsequently validated in section \ref{Model validation}, where computational results demonstrate the improved predictive capability.
Finally, section \ref{Summary and discussion} presents discussions and summarizes the contributions of the present study.

\section{Methodology }\label{Sec:Methodology}

\subsection{Governing equations and baseline model}\label{Sec:Governing equations and baseline model}
For mixing flows, the multicomponent RANS equations are solved considering  molecular transport and thermodynamic coefficients. 
The transport equations for the mean density $\bar{\rho}$, velocity $\tilde{u}_{i}$, total energy $\tilde{E}$ of the mixture, and mass fraction $\tilde{Y}_{\alpha}$ of species $\alpha$ are given as follows:
\begin{eqnarray}
&&\frac{\partial \bar{\rho }}{\partial t}+\frac{\partial \bar{\rho }{{{\tilde{u}}}_{j}}}{\partial {{x}_{j}}}=0,  \label{mass} \\ 
&&\frac{\partial \bar{\rho }{{{\tilde{u}}}_{i}}}{\partial t}+\frac{\partial \bar{\rho }{{{\tilde{u}}}_{i}}{{{\tilde{u}}}_{j}}}{\partial {{x}_{j}}}+\frac{\partial \bar{p}}{\partial {{x}_{i}}}-\bar{\rho}g_{i}=-\frac{\partial \tau_{ij}}{\partial {{x}_{j}}}+   
\frac{\partial {{{\bar{\sigma}}}_{ij}}}{\partial {{x}_{j}}},  \\
&&\frac{\partial \bar{\rho }\tilde{E}}{\partial t}+\frac{\partial (\bar{\rho }\tilde{E}+\bar{p}){{{\tilde{u}}}_{j}}}{\partial {{x}_{j}}}-\bar{\rho}\tilde{u}_{i}g_{i}=D_{E}+D_{K}+\frac{\partial }{\partial {{x}_{j}}}\left(-{\tau}_{ij}{{\tilde{u}}_{i}}+{{\bar{\sigma }}_{ij}}{{\tilde{u}}_{i}}   
 -\bar{q}_{c}-\bar{q}_{d}
\right),   \\
&&\frac{\partial \bar{\rho }{{{\tilde{Y}_{\alpha}}}}}{\partial t}+\frac{\partial \bar{\rho }{{{\tilde{Y}_{\alpha}}}}{{{\tilde{u}}}_{j}}}{\partial {{x}_{j}}}=D_{Y}+\frac{\partial }{\partial {{x}_{j}}}\left(\bar{\rho}\bar{D}\frac{\partial\tilde{Y}_{\alpha}}{\partial x_{j}}\right). \label{species}
\end{eqnarray}
The overbar and tilde represent the Reynolds and Favre averaged fields, respectively.
The subscripts $i,j=1,2,3$ denote  three spatial directions ($x,y,z$).
The $\bar{p}$ denotes pressure of the mean field.
The heat flux $\bar{q}_{c}$, the interspecies diffusional heat flux $\bar{q}_{d}$, and the viscous stress tensor $\bar{\sigma}_{ij}$ are given as $\bar{q}_{c}=-\bar{\kappa} \partial \tilde{T}/\partial x_{j}$, $\bar{q}_{d}=-\sum \bar{\rho} \bar{D}\bar{c}_{p,\alpha}\tilde{T}\partial \tilde{Y}_{\alpha}/\partial x_{j}$, $\bar{\sigma}_{ij}= 2\bar{\mu}(\tilde{S}_{ij}-\tilde{S}_{kk}\delta_{ij}/3)$, $\tilde{S}_{ij}=(\partial \tilde{u}_{i}/\partial x_{j}+\partial \tilde{u}_{j}/\partial x_{i})/2$.
Here $\bar{\mu}$, $\bar{D}$,  $\bar{\kappa}$, and $\bar{c}_{p,\alpha}$ represent dynamic viscosity, mass diffusivity, thermal conductivity,  and constant-pressure specific heat of species  $\alpha$, respectively. 
In correspondence with \citet{Livescu2013Num}, the mixture's EOS $\bar{p}\tilde{M}=\bar{\rho}R\tilde{T}$ is used, which is calculated with the assumptions of iso-temperature (i.e. $T_1=T_2=...=T_{\alpha}$) and partial-pressure (i.e. $p=\sum {p}_{\alpha}$), and  the fluid properties of the mixture are determined using the species-linearly weighted assumption.
  
Equations (\ref{mass})$\sim$(\ref{species}) are deduced based  on the ensemble average operation, which results in the unclosed terms, i.e. the Reynolds stress $\tau_{ij}$, the turbulent diffusion terms $D_{E}$, $D_{K}$ and $D_{Y}$ of the total energy, the turbulent kinetic energy (TKE, $\tilde{K}$), and the mass fraction, respectively.
With the widely used gradient diffusion assumption (GDA), that the turbulent flux  of an arbitrary physical variable $f$ can be proportional to its mean gradient, i.e. 
\begin{equation}  \label{GDA}
-\bar{\rho}\widetilde{u_{j}^{''}f^{''}}=\frac{\mu_{t}}{N_{f}}\frac{\partial \tilde{f}}{\partial x_{j}} \quad or \quad
-\overline{\rho u_{j}^{'}f^{'}}=\frac{\mu_{t}}{N_{f}}\frac{\partial \bar{f}}{\partial x_{j}}, 
\end{equation}
where `$'$' and  `$''$' denote the Reynolds and Favre fluctuations respectively, as well as  $N_{f}$ being a model coefficient, these turbulent diffusion terms are modeled as $ {D}_{E}=\frac{\partial}{\partial x_{j}}\left(\frac{\mu_{t}}{N_{h}}\frac{\partial \tilde{h}}{\partial x_{j}}\right )$, ${D}_{K}=\frac{\partial}{\partial x_{j}}\left(\frac{\mu_{t}}{N_{K}}\frac{\partial \tilde{K}}{\partial x_{j}}\right )$, ${D}_{Y}= \frac{\partial}{\partial x_{j}}\left(\frac{\mu_{t}}{N_{Y}}\frac{\partial \tilde{Y}}{\partial x_{j}}\right)$, respectively, the enthalpy $h=e+p/\rho=c_pT$, $e$ representing internal energy. 

In the $K-L-\gamma$ model, the TKE $\tilde{K}$ and the turbulent length scale $\tilde{L}$ are used to describe the turbulent viscosity $\nu_{t}$, which is limited by the intermittent factor $\gamma$ to enable this model to describe mixing transition flows \citep{Xie2025intermittency}.
The $\mu_{t}$ is expressed as $\mu_{t}=\bar{\rho}\nu_t=C_{\mu}\bar{\rho} \gamma\tilde{L}\sqrt{2\tilde{K}}$, where $C_{\mu}$ is a model coefficient.
Based on the classical Boussinesq eddy viscosity hypothesis, the Reynolds stress is modeled as $ {{{\tau }}_{ij}}={{C}_{P}}\bar{\rho }\tilde{K}{{\delta }_{ij}}-2{{\mu }_{t}}(\tilde{S}_{ij}-\tilde{S}_{kk}{{\delta }_{ij}}/3)$,  here $C_{P}$ is a  model coefficient and $\delta_{ij}$ is the Kronecker symbol. 
To describe the spatio-temporal evolutions of the $\tilde{K}$, $\tilde{L}$ and $\gamma$, their transport equations are introduced as follows:
\begin{eqnarray}
\label{eq:trans_TKE}
&&\frac{\partial \bar{\rho }\tilde{K}}{\partial t}+\frac{\partial \bar{\rho }\tilde{K}{{{\tilde{u}}}_{j}}}{\partial {{x}_{j}}}=-{{{\tau }}_{ij}}\frac{\partial {{{\tilde{u}}}_{i}}}{\partial {{x}_{j}}}+\frac{\partial }{\partial {{x}_{j}}}\left(\frac{{{\mu }_{t}}}{{{N}_{K}}}\frac{\partial \tilde{K}}{\partial {{x}_{j}}}\right)+{{S}_{Kf}}-{{C}_{D}}\bar{\rho }{{\left( \sqrt{2\tilde{K}}\right)}^{3}}/{\tilde{L}},  \\
\label{eq:trans_L}
&&\frac{\partial \bar{\rho }\tilde{L}}{\partial t}+\frac{\partial \bar{\rho }\tilde{L}{{{\tilde{u}}}_{j}}}{\partial {{x}_{j}}}=\frac{\partial }{\partial {{x}_{j}}}\left(\frac{{{\mu }_{t}}}{{{N}_{L}}}\frac{\partial \tilde{L}}{\partial {{x}_{j}}}\right)+{{C}_{L}}\bar{\rho }\sqrt{2\tilde{K}}+{{C}_{C}}\bar{\rho }\tilde{L}\frac{\partial {{{\tilde{u}}}_{j}}}{\partial {{x}_{j}}}, \\
&&\frac{\partial \bar{\rho }\gamma}{\partial t}+\frac{\partial \bar{\rho }\gamma{{{\tilde{u}}}_{j}}}{\partial {{x}_{j}}}=\frac{\partial }{\partial {{x}_{j}}}\left(\frac{{{\mu }_{t}}}{{{N}_{\gamma}}}\frac{\partial \gamma}{\partial {{x}_{j}}}\right)+P_{\gamma}-\epsilon.
\end{eqnarray}
The $P_{\gamma}=F_{onset}G_r$ is the production term of the $\gamma$ equation,  the dissipation term $\epsilon=\gamma P_{\gamma}$ ensuring the intermittent factor varying from 0 to 1.
The $F_{onset}$ serves as a transition switch, and $G_r$ describes the growth of $\gamma$, which are expressed as 
\begin{equation}\label{fonset} 
F_{onset}=1-\frac{1}{e^{max(0,Re_{t}-Re_{tra})}}, \quad Re_{t}=\frac{\tilde{L}\sqrt{\tilde{K}}}{\nu},
\end{equation}
\begin{equation} \label{Gr}
G_{r}=C_{1}\sqrt{\gamma}\left\{ 
(1-\gamma) \left[\bar{\rho}\sqrt{2\tilde{S}_{ij}\tilde{S}_{ij}}+C_{2}\sqrt{max \left(0,-\frac{\partial \bar{\rho}}{\partial x_k} \frac{\partial \bar{p}}{\partial x_k}\right)}\right]+ \left(\gamma S_{Kf}-{{{\tau }}_{ij}}\frac{\partial {{{\tilde{u}}}_{i}}}{\partial {{x}_{j}}}\right)/\tilde{K}
\right\}. 
\end{equation}
Here, $C_1$ and $C_2$ are model coefficients, $\nu$ denoting the kinetic viscosity, and $Re_{tra}$ denoting a pre-defined key transition Reynolds number set to 10, following \citet{Xie2025intermittency}.

The buoyancy-driven source term ${{S}_{Kf}}$ is modeled based on the following advanced formation \citep{Xiao2020Modeling},
\begin{equation}
\label{Skf}
S_{Kf} = 
\begin{cases} 
- C_{B} \frac{\partial \bar{p}}{\partial x_{i}} \sqrt{2\tilde{K}} \left| (1 - \omega_{l}) A_{0i} + \omega_{l} A_{ssi} \right|,  &if   \Theta_{gf}< \Theta_{gm}, \\
- C_{B} \frac{\partial \bar{p}}{\partial x_{i}} \sqrt{2\tilde{K}} \max \left[0, (1 - \omega_{l}) A_{0i} + \omega_{l} A_{ssi}\right],  & otherwise,
\end{cases}
\end{equation}
where $C_B$ is a model coefficient.
The mean- and turbulent- field accelerations $\Theta_{gm}$ and $\Theta_{gf}$  are used to identify RM and RT mixings, whose  expressions can be seen in \citet{Xiao2020Modeling}.
The initial cell Atwood number ${{A}_{0i}}$ describes the density's discontinuous characteristic at the interfaces between fluids, and can be expressed by the reconstructed values of the densities at the cell’s faces \citep{Kokkinakis2015Two}.  
The local Atwood number ${{A}_{ssi}}$ in the self-similar regime depicts the density's gradual variation, and is calculated by 
\begin{equation} \label{Assi}
  {{A}_{ssi}}=\frac{{{C}_{A}}{\tilde{L}}}{\bar{\rho }+\tilde{L}\left| \frac{\partial \bar{\rho }}{\partial x_{i}} \right|}\frac{\partial \bar{\rho }}{\partial x_{i}}, 
\end{equation}
where $C_A$ is a model coefficient.
The weighting factor $\omega_{l}=min(\frac{\tilde{L}}{\Delta x},1)$ ($\Delta x$ is the grid scale) ensures that mixing evolution is solely decided by $A_{ssi}$ when $\tilde{L}$ becomes sufficiently large, while initially is dominated by $A_{0i}$ \citep{Kokkinakis2015Two}.

Equations (\ref{mass})$\sim$(\ref{species}), (\ref{eq:trans_TKE})$\sim$(\ref{Assi}) constitute  the complete $K-L-\gamma$ mixing transition model, which has demonstrated a good  capability for predicting incompressible RT/RM mixing transition flows.
More details about this model can be found in \citet{Xie2025intermittency}. 
Given the prominent transition processes observed in density-stratified compressible RT mixing flows, the $K-L-\gamma$ model is  adopted as the baseline model in this study to be expected to capture the transition effect.
In the next section, performance of the baseline model for the targeted flows will be evaluated.
In addition, the widely used $K-L$ model, which is designed for fully developed turbulent mixing flows, is also tested for comparison, highlighting the necessity of considering the transitional effects by introducing the $\gamma$ equation.

\subsection{Baseline model evaluation for compressibility effects}\label{Convective instability in density-stratified compressible RT flow}
The evaluated cases originate from the three-dimensional (3D) simulation of initially isothermal density-stratified compressible RT mixing implemented by \citet{Luo2022Mixing}.
These simulations solved the multicomponent Navier-Stokes equations for  cases with different density stratification parameters $S_r=$ 0.5, 1, 2 and 3.
An eighth-order central compact finite difference scheme were employed on a uniform grid of $1024\times512\times512$ grid points in a rectangular box $x\times y\times z$ of $2L_r\times L_r\times L_r$, $L_r$ = 1. 
Time integration is performed using a third-order Runge-Kutta scheme.
The simulations of \citet{Luo2022Mixing} are thus highly resolved and referred to as HiFi simulations in the present study.

\begin{figure}
\centering
\subfigure{
\includegraphics[width=0.45\textwidth]{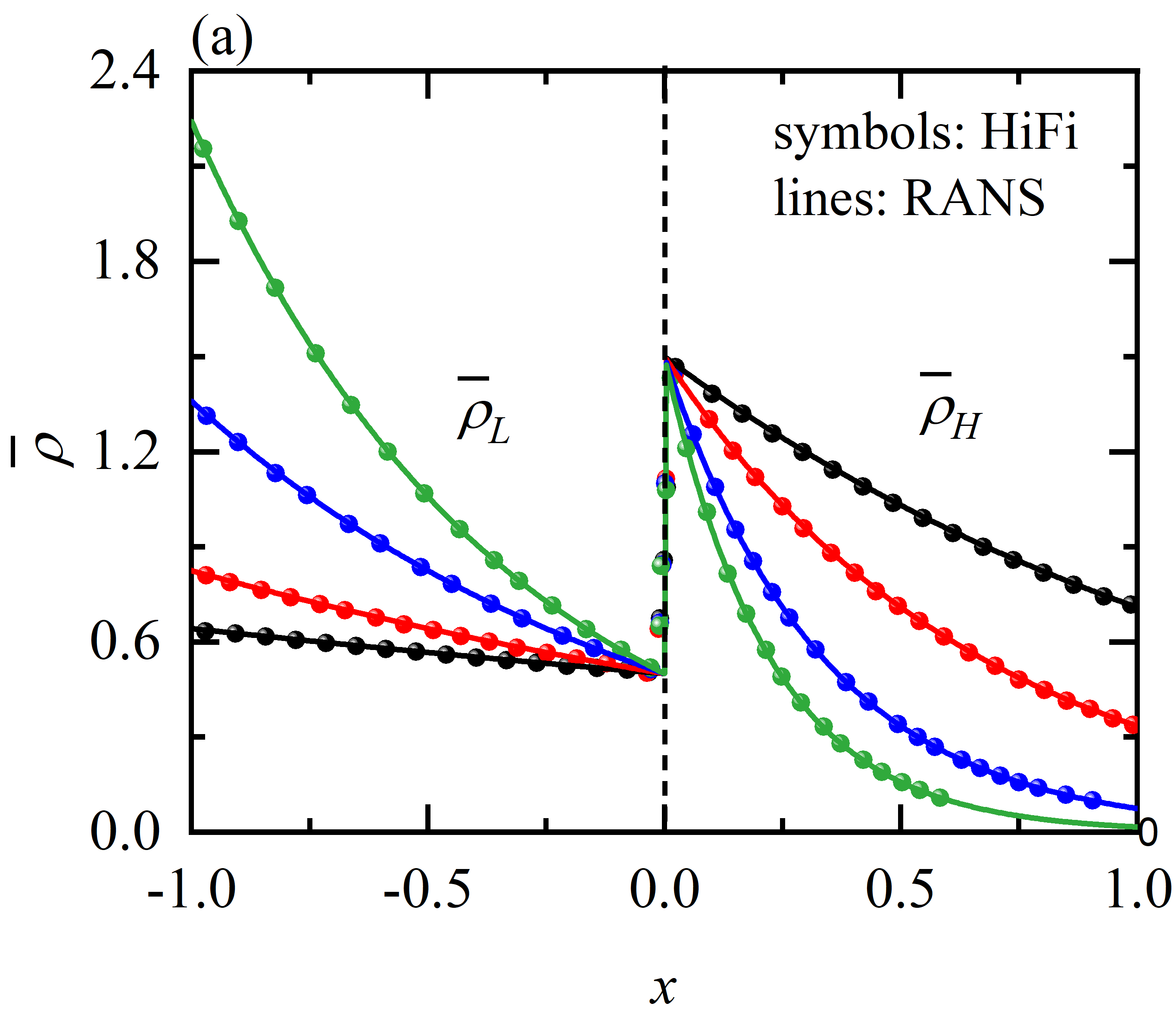}}
\subfigure{
\includegraphics[width=0.45\textwidth]{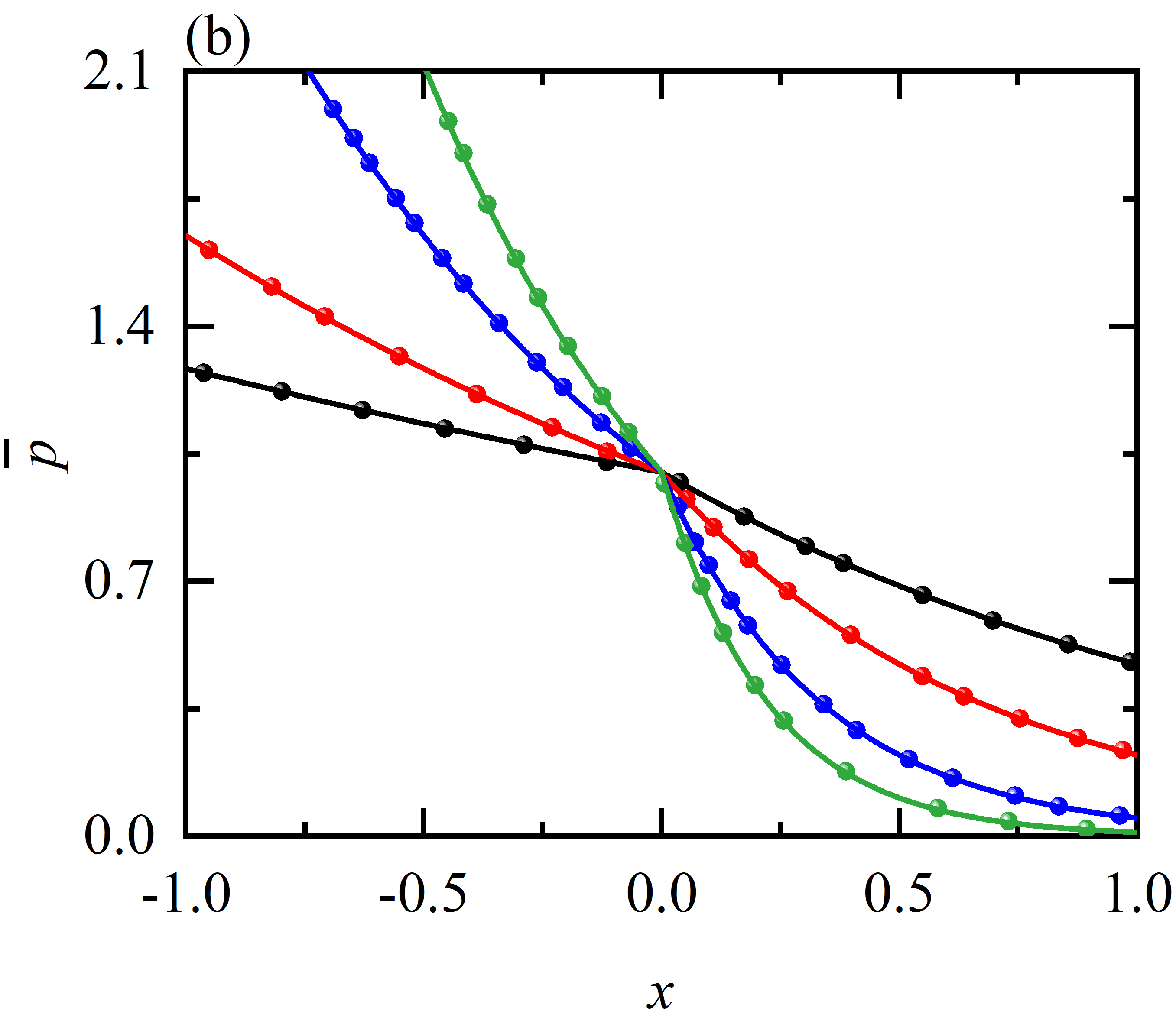}}
\caption{Initial distributions for density and pressure fields.
The symbols and lines represent the initial flow fields of HiFi simulations and RANS simulations, respectively. 
The lines colored by black, red, blue, and green correspond to cases of $S_r=$ 0.5, 1, 2 and 3, respectively.
} \label{ini_field}
\end{figure}

Due to homogeneity of the flow in the spanwise plane (i.e. the $y-z$ plane), the 3D cases are statistically reduced  to the one-dimensional (1D) equivalents for RANS computations. 
The flow configuration is described as follows.
The light and heavy fluids, represented by the subscript $L$ and $H$ respectively, are initially placed in the computational domains $[-L_x,0] cm$  and $[0,L_x]cm$  respectively, with $L_x=L_r=1$.
A constant gravitational acceleration $g=1 cms^{-2}$ is imposed along the $-x$ direction. 
Both fluids  are compressible ideal gases with the same adiabatic exponent $\Gamma=5/3$, and the molar mass $M_L$ and $M_H$ are $0.029 kg/mol$ and $0.087 kg/mol$, respectively, yielding an Atwood number $A_t\equiv\frac{M_H-M_L}{M_H+M_L}=0.5$, consistent with the definition in \citet{Luo2022Mixing}.
The stratification parameter $S_r$ is defined based on the formula (\ref{Sr}).
The characteristic Reynolds number $Re\equiv\frac{g^{1/2}{L_r}^{3/2}}{\mu_r/\rho_r}=10000$, the Prandtl number and the Schmidt number are   0.7 and 1.0, respectively.
Reference quantities used to define these dimensionless parameter include the temperature $T_r=T_H=T_L$,  molar mass $M_r=(M_H+M_L)/2$, dynamic viscosity $\mu_r=\mu_H=\mu_L$,  density $\rho_r=(\rho_{H,0}+\rho_{L,0})/2$, where $\rho_{H,0}$ and $\rho_{L,0}$ represent the initial densities on either side of the interface.
Additionally, the characteristic time scale is defined as $\tau=\sqrt{L_r/(A_tg)}$.
All reference quantities are consistent with the HiFi simulations to ensure direct comparability between RANS and HiFi results.
The system is initialized with a hydrostatic background state in thermal equilibrium.
By setting  $\tilde{u}=0$ in the momentum equation, we can  obtain the initial    density and pressure distributions:
\begin{eqnarray}
    &&p_m=p_Iexp\left(-\frac{gx}{(R/M_m)T_I}\right), \\
    &&\rho_m=\frac{p_I}{(R/M_m)T_I}exp\left(-\frac{gx}{(R/M_m)T_I}\right),
\end{eqnarray}
where $p_I$ and $T_I$ represent the initial  pressure  and temperature at the interface, respectively, the subscript $m$ representing either heavy ($H$) or light ($L$) fluid.
Specifically, $p_I=1 g cm^{-1}s^{-2}$, $T_I$ equals to $T_r$  determined based on the formula (\ref{Sr}).
Figure \ref{ini_field} shows  comparisons of the initial density and pressure fields from RANS and HiFi simulations, ensuring the correctness of  the present configurations. 

Performance of the baseline model for the targeted flows is tested here.
The RANS calculations are implemented in the finite difference code for compressible fluid dynamics developed by Y. Zhang \emph{et al.} since 2016, which has been extensively validated for RT and RM mixing studies \citep{Xie2025intermittency,zhang2020Methodology,Xiao2025Local,Li2019On,Xiao2022exp,Xie2023Data}.
A combination of MUSCL5, i.e. the improved total variation diminishing
scheme with a fifth-order limiter, and the Riemann solver proposed
by \citet{Toro1994RestorationOT} is used to calculate the convection terms.
To reduce the numerical dissipation, low Mach number modification \citep{Thornber2008on,Thornber2008An} is used during the reconstruction of the half-point flux. 
The diffusion terms are calculated with the sixth-order central difference scheme. 
The temporal integration marches with a third-order Runge–Kutta scheme, with Courant-Friedrichs-Lewy number 0.05.

The  model coefficients used in the baseline model are listed in table \ref{tabcoe}, which are determined based on the methodology proposed by \citet{zhang2020Methodology}.
For initialization of the turbulence model variables, we adopt the common practices: nearby the interface the initial turbulent length scale $\tilde{L}_0$ is $0.04cm$, approximating to the corresponding 3D simulation's initial perturbation width, the initial TKE  $\tilde{K}_0$ is $0.0001A_tg\tilde{L}_0(cm/s)^2$, as well as the initial intermittent factor $\gamma_0=1\times10^{-16}$ following \citet{Xie2025intermittency}. 
\begin{table}
\caption{\label{tabcoe} Model coefficients of the baseline model.}
\centering
\begin{threeparttable}
\setlength{\tabcolsep}{1mm}{
\begin{tabular}{llllllllllllll}
\toprule
$C_C$ & $C_P$ &  $C_L$ & $N_K$ &  $N_L$ &  $N_\gamma$ &  $N_Y$ & $N_h$ &  $C_\mu$ & $C_B$ & $C_A$ & $C_D$  & $C_1$ & $C_2$\\
\midrule
 1/3 & 2/3  &  0.189
& 1.069  & 0.096  & 0.100 & 0.869 & 0.869 & 1.186 & 0.482 & 7.088  & 0.200 & 2.200 & 0.015
\\
\bottomrule
\end{tabular}}
\end{threeparttable}
\end{table}

\begin{figure}
\centering
\subfigure{
\includegraphics[width=0.45\textwidth]{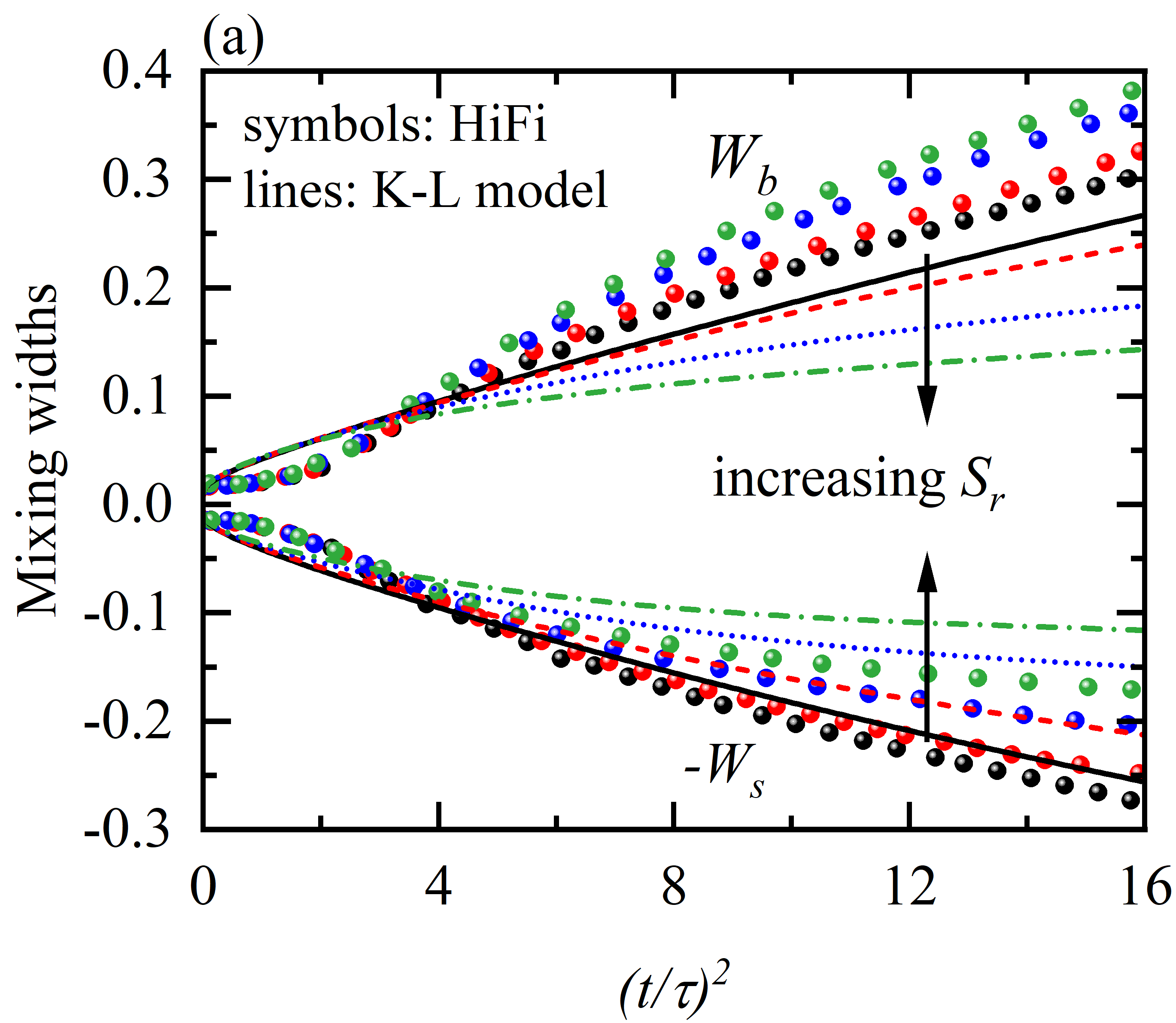}}
\subfigure{
\includegraphics[width=0.45\textwidth]{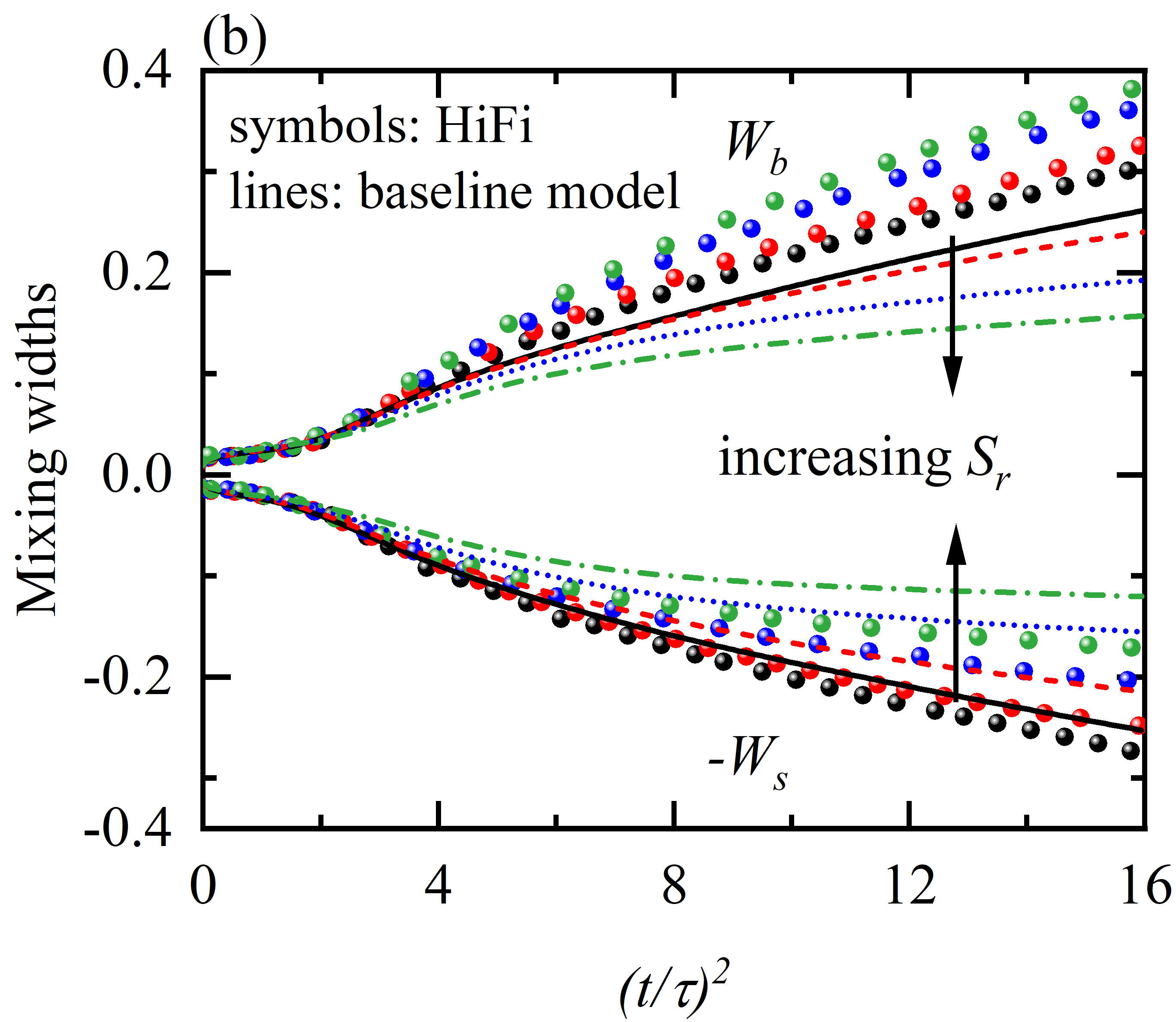}}
\caption{Temporal evolution of the  mixing width predicted by (a) the $K-L$ model and (b) the $K-L-\gamma$ baseline model. 
The colors  used to distinct different $S_r$ cases are consistent with  figure \ref{ini_field}.
} \label{evalution}
\end{figure}

Figure \ref{evalution} presents  temporal evolution of the mixing width, an important statistical quantity used to characterize the global evolution of mixing flows.
The mixing width is defined as:
\begin{equation}\label{dewid}
    W_b=3\int_0^{L_x}\bar{X}(1-\bar{X})dx, \quad W_s=3\int_{-L_x}^{0}\bar{X}(1-\bar{X})dx. 
\end{equation}
The mole fraction $X$  is calculated by the mass fraction \citep{Luo2022Mixing}.
The $W_b$ and  $W_s$ represent the bubble mixing width, i.e. distance of the light fluid penetrating the heavy fluid, and the spike mixing width, i.e. distance of the heavy fluid penetrating the light fluid, respectively.
Figure \ref{evalution}(a) shows the prediction results of the $K-L$ model, which gives a  $t^2$-scaling law throughout the entire simulation for all cases, failing to depict the transition process.
This is expected because the $K-L$ model is only suitable for the incompressible fully developed turbulent mixing flows. 
In contrast, the baseline model performs well during the instability development and mixing transition stages at approximately $(t/\tau)^{2}<4$, providing predictions consistent with HiFi data for all cases, as illustrated in figure \ref{evalution}(b). 
It is attributed to that the baseline model can capture mixing transition processes effectively. 
Beyond the transition stage, the mixing width predicted by the baseline model exhibits a quadratic growth in time, according with the late-stage self-similar evolution of incompressible RT mixing flows.
Figure \ref{evalution} highlights the superior capability of the baseline model in predicting mixing transition flows.

Notably, during the early stage, the baseline model gives  consistent predictions with HiFi simulations across all compressibility strengths considered here, even though it is essentially an incompressible model.                
It it attributed  to the negligible compressibility effects in early development of the density-stratified RT mixing flows, where behavior closely resembles incompressible cases \citep{Gauthier2017Compressible}. 
However, the baseline model's performance deteriorates significantly during intermediate and late stages.
The discrepancies between RANS predictions and HiFi data progressively widen with increasing $S_r$ values.
It can be understood as follows. 
The $S_r$ directly correlates with the compressibility strength: a smaller $S_r$ corresponds to a weaker compressibility effect, flow approaching to the incompressible case as $S_r\rightarrow0$. 
In the case of $S_r=0.5$, the  compressibility is not prominent so that the baseline model can give a comparatively reasonable  prediction.  
For the cases with larger $S_r$, strong density stratification brings strong compressibility effect, resulting in failures of the baseline model.
The degradation in predictive accuracy directly correlates with the strength of density stratification, highlighting the need for compressibility corrections in the existing RANS modeling.

\begin{figure} 
\centering
\subfigure{
\includegraphics[width=0.32\textwidth]{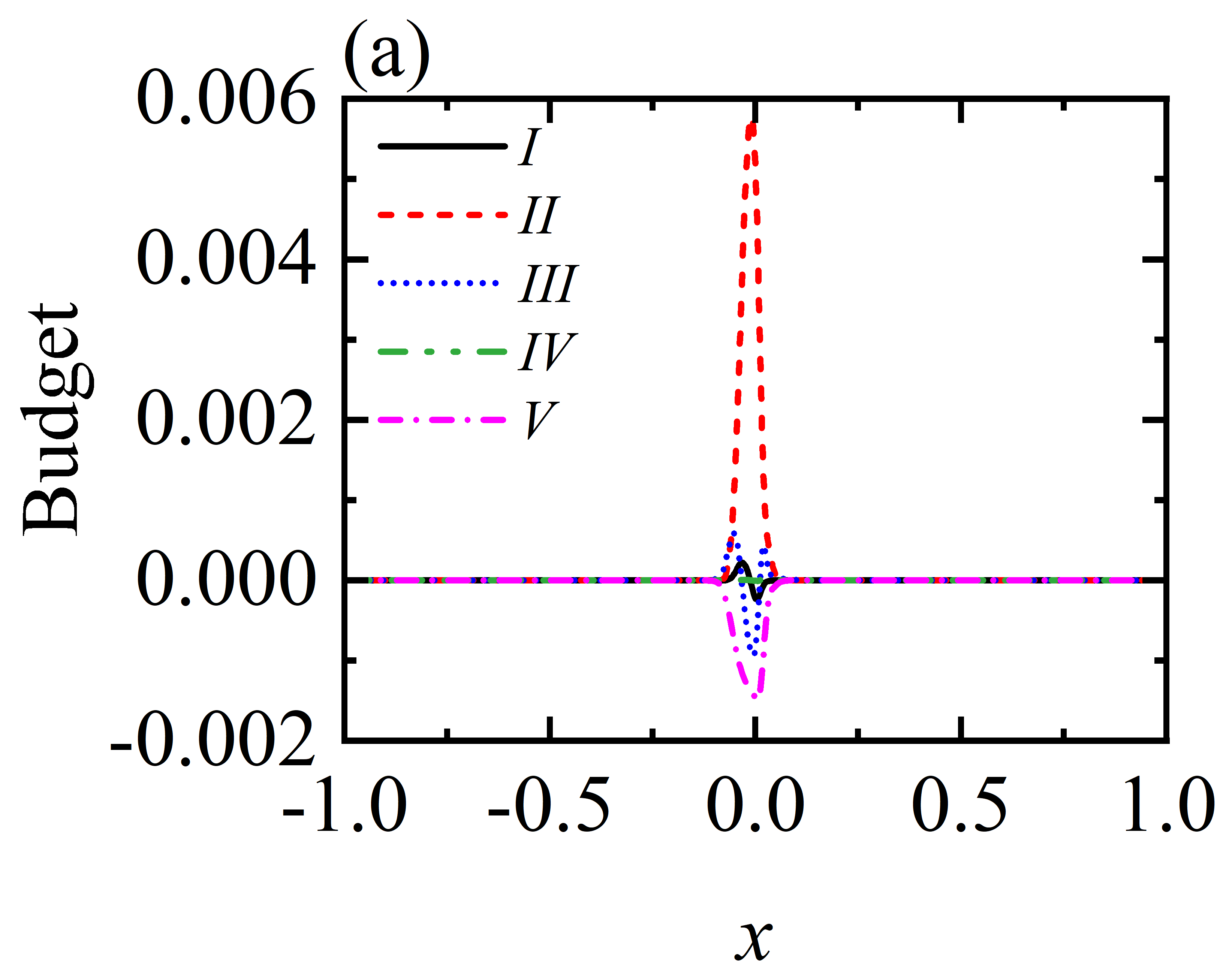}}
\subfigure{
\includegraphics[width=0.32\textwidth]{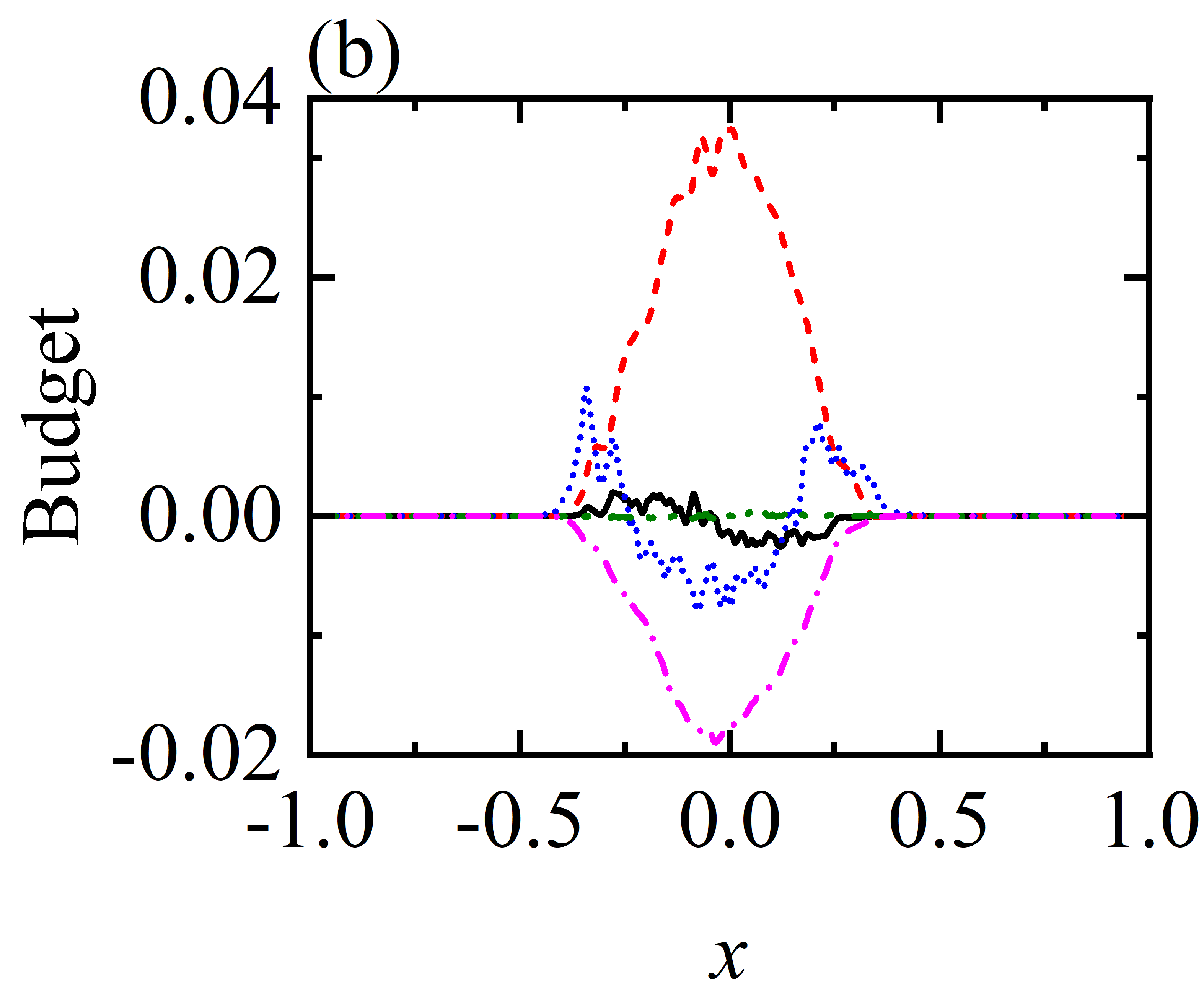}} 
\subfigure{
\includegraphics[width=0.32\textwidth]{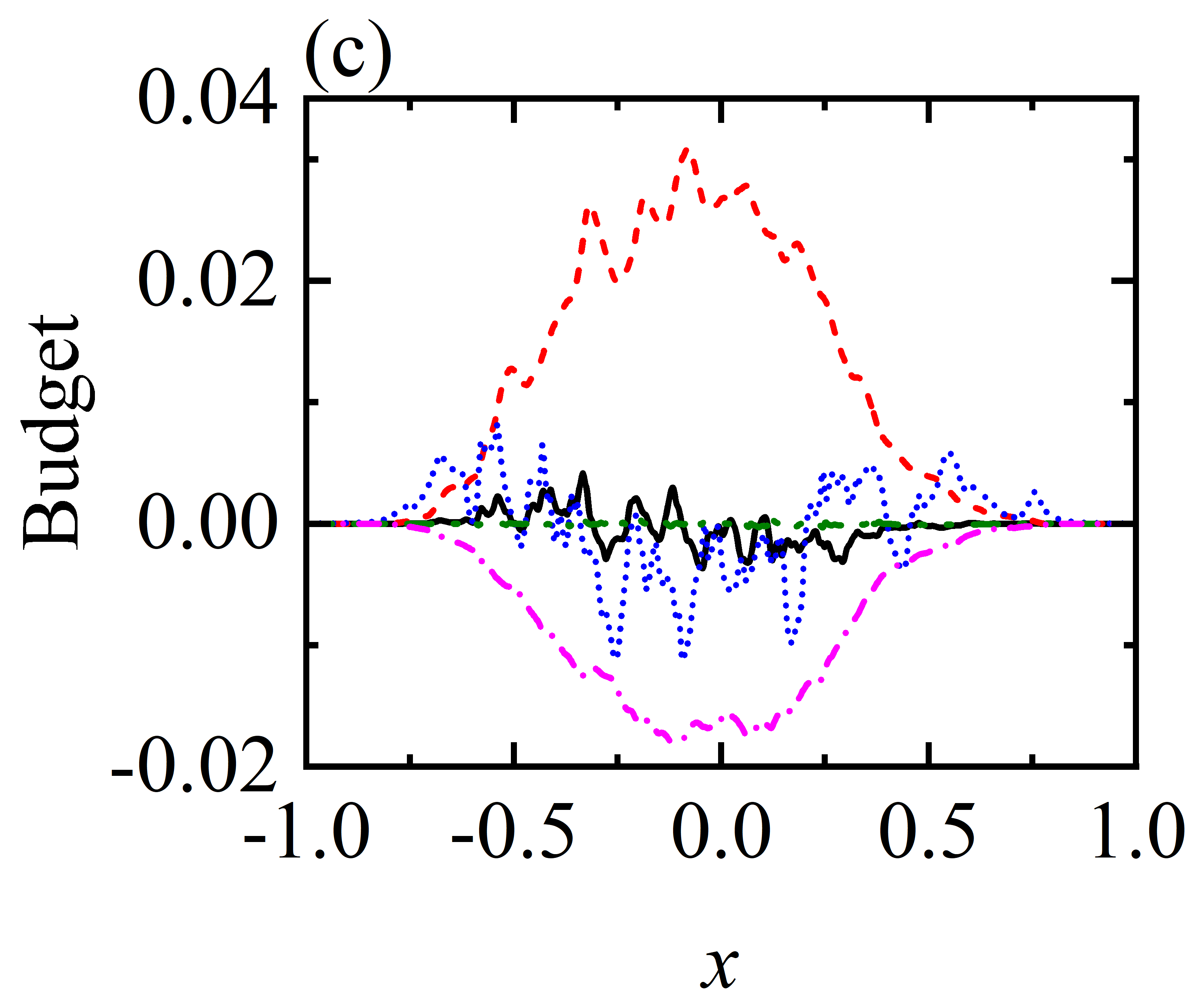}}
\subfigure{
\includegraphics[width=0.32\textwidth]{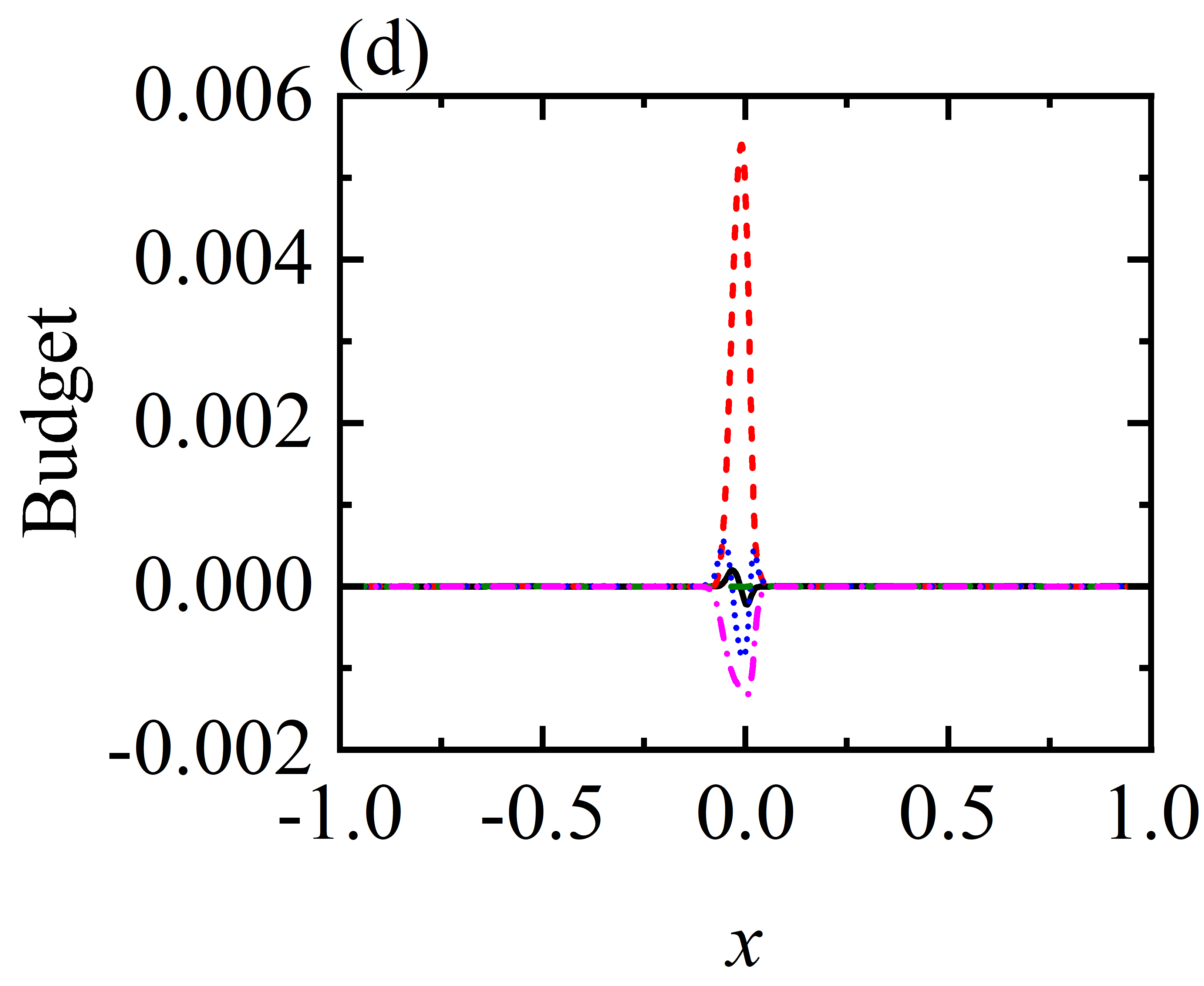}}
\subfigure{
\includegraphics[width=0.32\textwidth]{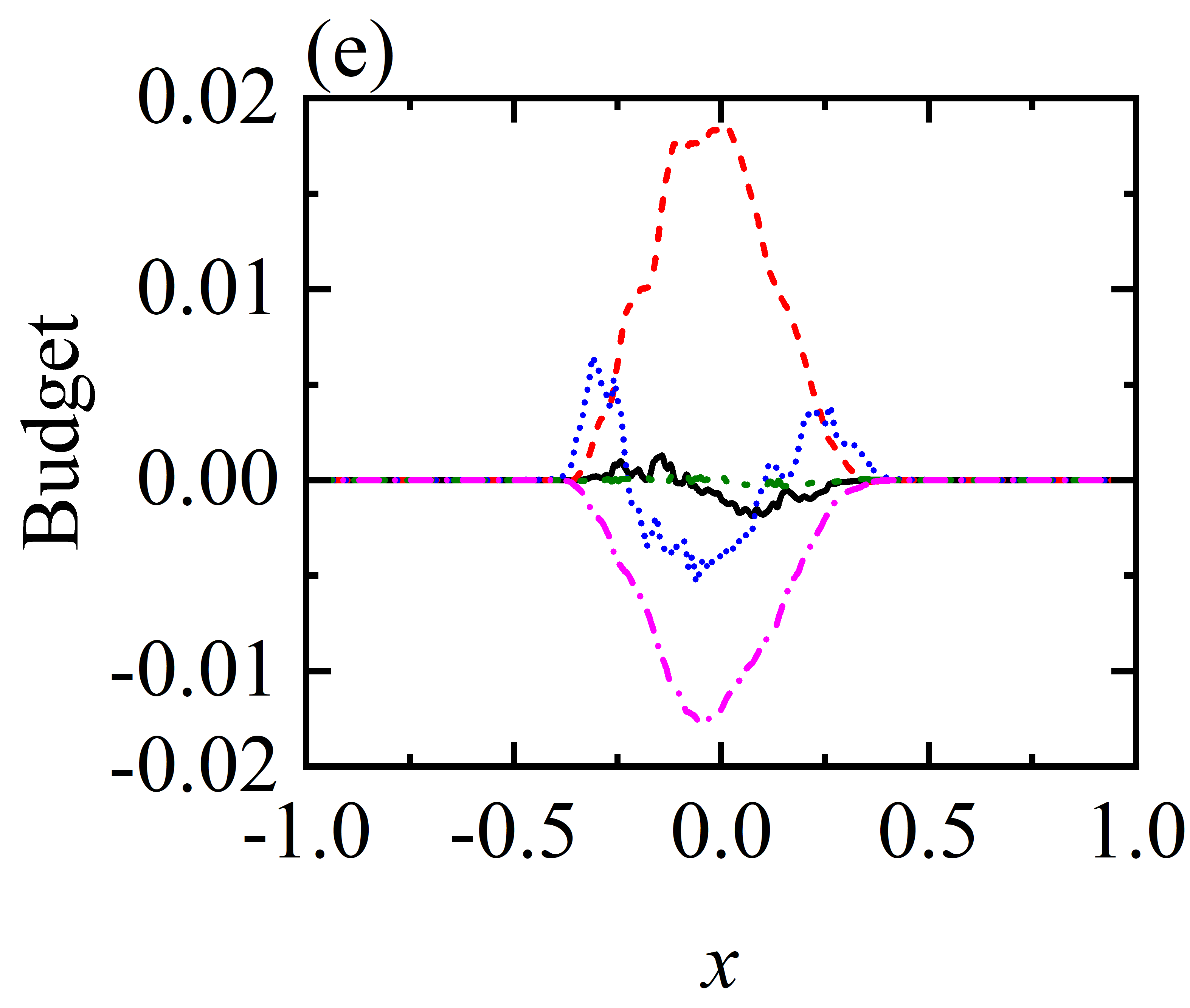}}
\subfigure{
\includegraphics[width=0.32\textwidth]{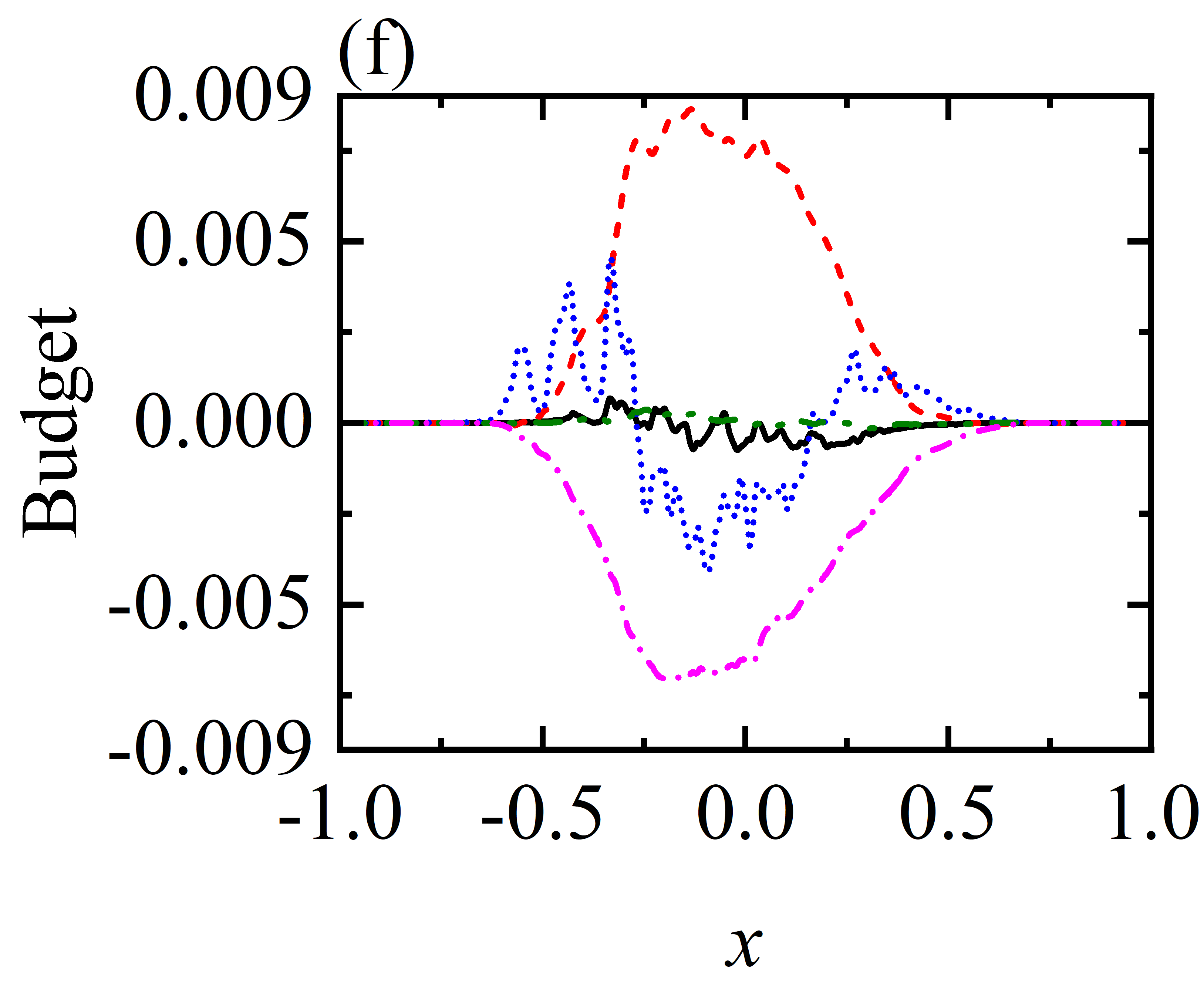}}
\caption{Budgets for terms $I\sim V$ of the TKE equation (\ref{oriTKE}). Subfigures (a)$\sim$(c) and (d)$\sim$(f) correspond to moments of $t/\tau=$1, 2.5 and 4 for cases of $S_r=$ 0.5 and 3 respectively.} 
\label{budget}
\end{figure}

The baseline model consistently  underestimates  the evolution of mixing width across all cases examined here.
To understand the discrepancy, the underlying modeling philosophy is carefully analyzed.
For the $K-L-\gamma$  and the $K-L$ models, the TKE transport equation  plays a crucial role in governing the dynamic evolution of both mixing and turbulence. 
Following \citet{Chassaing2002Variable} and \citet{Gauthier2017Compressible}, the exact  averaged  TKE equation without any modeling is written as:
\begin{equation} \label{oriTKE}
\frac{\partial(\bar{\rho}\tilde{K})}{\partial t}+\frac{\partial(\bar{\rho}\tilde{K}\tilde{u}_j)}{\partial x_j}=\underbrace{-\overline{\rho u^{''}_iu^{''}_j}\frac{\partial \tilde{u_i}}{\partial x_j}}_{I}\underbrace{-\overline{u^{''}_i}\frac{\partial \bar{p}}{\partial x_i}}_{II}\underbrace{-\frac{\partial \left(\overline{\rho u^{''}_jK}+\overline{ p^{'}u^{''}_j}+\overline{ \sigma_{ij}u^{''}_i}\right)}{\partial x_j}}_{III}\underbrace{+\overline{p^{'}\frac{u^{''}_i}{\partial x_i}}}_{IV}\underbrace{-\overline{\sigma_{ij}\frac{\partial u^{''}_i}{\partial x_j}}}_{V},
\end{equation}
where terms $I\sim V$ represent shear production, turbulent mass flux (buoyancy production), diffusion, pressure-strain correlation, and dissipation, respectively.
Based on the HiFi simulations, budgets of the terms $I\sim V$ at $t/\tau=$1, 2.5, and 4 are given in figure \ref{budget}, where the cases of $S_r=$ 0.5 and 3 are chosen to represent  the weak and strong density stratification cases.
Figure \ref{budget} reveals  that the  important terms in right hand side of the TKE equation (\ref{oriTKE}) are the turbulent mass flux (term $II$) and dissipation (term $V$) terms. 
Given the dissipation enslaved by the production, the turbulent mass flux $-\overline{u^{''}_i}\frac{\partial \bar{p}}{\partial x_i}$ is thus considered as the dominant term.

In the $K-L-\gamma$ model, the turbulent mass flux $-\overline{u^{''}_i}\frac{\partial \bar{p}}{\partial x_i}$ is closed using the term $S_{Kf}$, as expressed by the combination of (\ref{Skf}) and (\ref{Assi}). 
Therefore, it is natural to analyze the modeling mechanism of the formulas (\ref{Skf}) and (\ref{Assi}), to explore  underlying reasons.
The key physical mechanism incorporated in   $S_{Kf}$ is the baroclinic term $\frac{\partial \bar{p}}{\partial x_{i}}\frac{\partial \bar{\rho}}{\partial x_{i}}$, which acts as an effective production source when $\frac{\partial \bar{p}}{\partial x_{i}}\frac{\partial \bar{\rho}}{\partial x_{i}}<0$; otherwise $S_{Kf}=0$.
The latter condition leads to a stagnation of mixing evolution, consistent with the observations in figure \ref{evalution}, where the growth rate of the mixing width approaches zero for cases with $S_r=$ 2 and 3 during late-stage development.
This behavior can be attributed to the premature disappearance of the baroclinic term in the present flows.
Specifically, due to the initial density stratification, the sign of the density gradient $\frac{\partial \bar{\rho}}{\partial x_{i}}$  transitions from positive to negative too early in the mixing region.
Since the pressure gradient  $\frac{\partial \bar{p}}{\partial x_{i}}$ remains negative under gravitational field, the product  $\frac{\partial \bar{p}}{\partial x_{i}}\frac{\partial \bar{\rho}}{\partial x_{i}}$ becomes positive once $\frac{\partial \bar{\rho}}{\partial x_{i}}<0$, forcing $S_{Kf}=0$. 
Consequently, the baseline model fails to supply the necessary source term to sustain mixing and turbulence evolution, thereby deviating from the true physical behavior.

Furthermore, stability analysis can serve as a practical tool for investigating compressible stratified flows characterized by finite density gradients and continuous profiles \citep{Gauthier2010Compressibility}.
Notably, local instability criteria can effectively identify unstable flow regions.
For incompressible flow, the necessary and sufficient condition of a stratified heterogeneous fluid being stable is $\frac{\partial\rho}{\partial x_i}<0$  everywhere, whereas for any region where  $\frac{\partial\rho}{\partial x_i}>0$ the stratification is considered unstable \citep{Chandrasekhar1961Hydrodynamic}.
In gravitational field, the instability condition is expressed as
\begin{equation} \label{Q1}
-\frac{\partial\rho}{\partial x_i}\frac{\partial p}{\partial x_i}>0.    
\end{equation}
For compressible flow, the instability criterion  within the linear approximation for the Euler equations is suggested as \citep{Gamalii1980Hydrodynamic}:
\begin{equation} \label{Q2}
    -\frac{\partial p}{\partial x_i}\left(\frac{\partial\rho}{\partial x_i}-\frac{1}{c^2}\frac{\partial p}{\partial x_i}\right)>0,
\end{equation} 
where $c$ is the isentropic speed of sound.
Obviously, the compressible criterion (\ref{Q2}) can restore to the incompressible one (\ref{Q1}) as $c\rightarrow\infty$.

\begin{figure} 
\centering
\subfigure{
\includegraphics[width=0.32\textwidth]{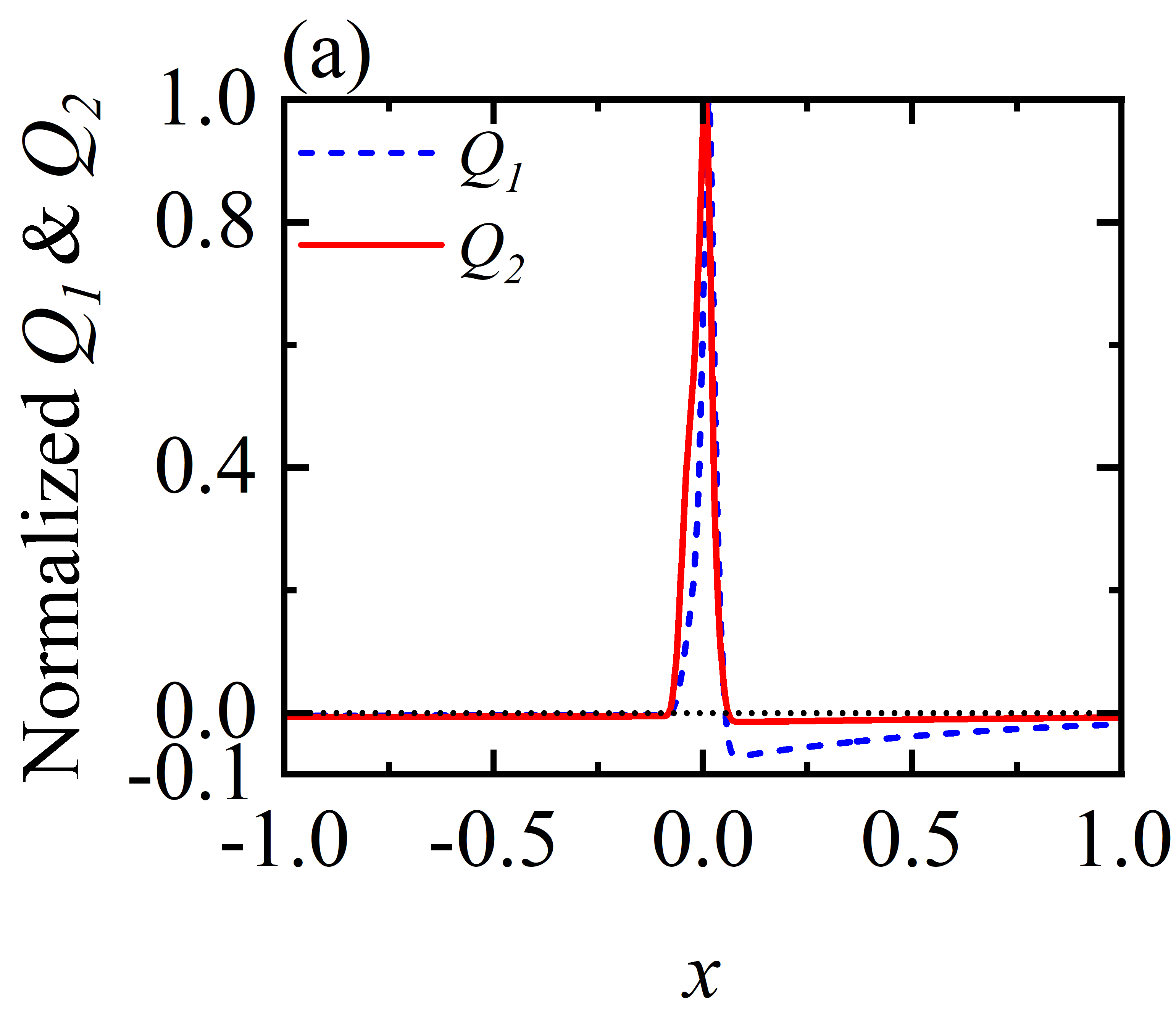}}
\subfigure{
\includegraphics[width=0.32\textwidth]{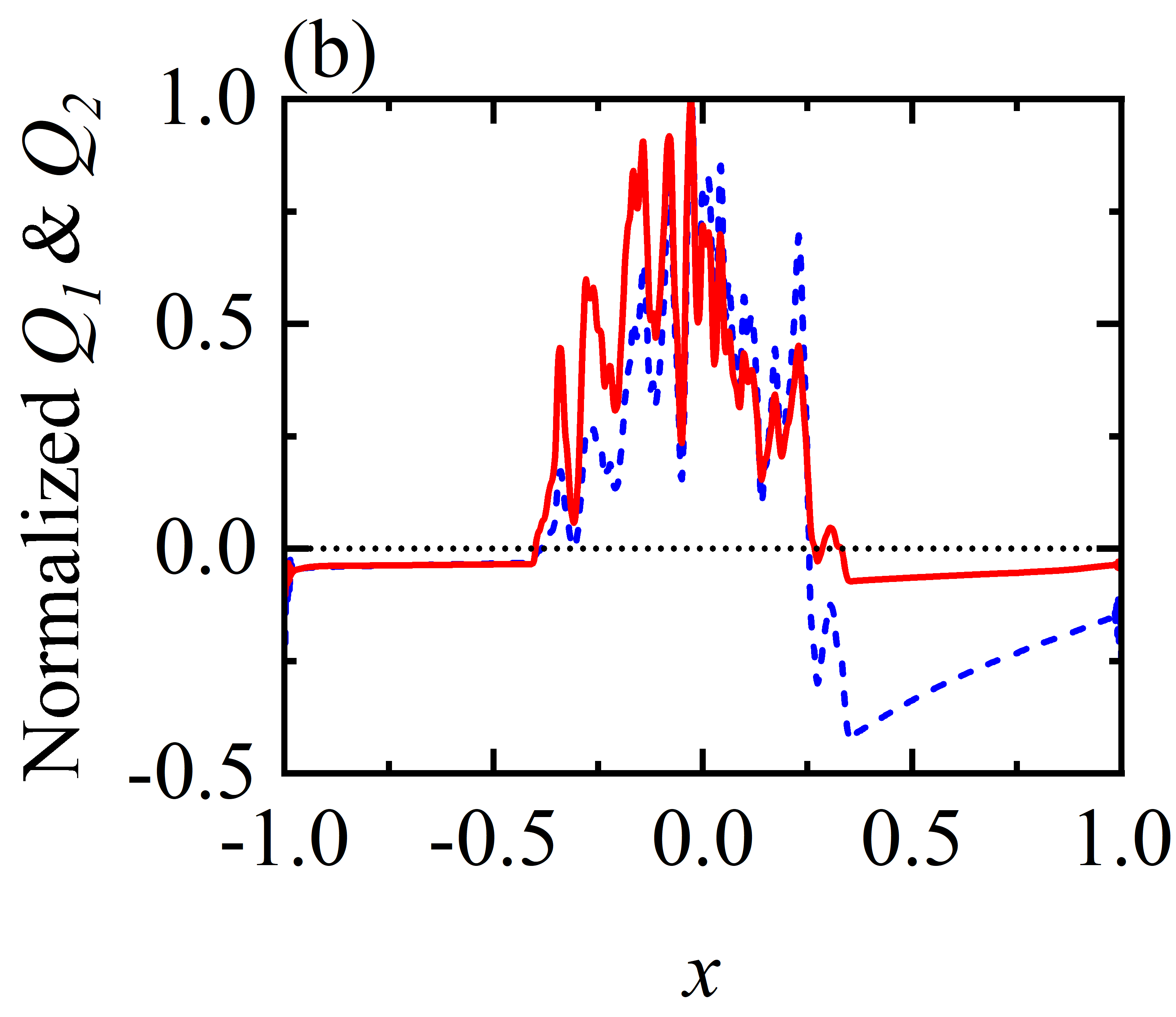}} 
\subfigure{
\includegraphics[width=0.32\textwidth]{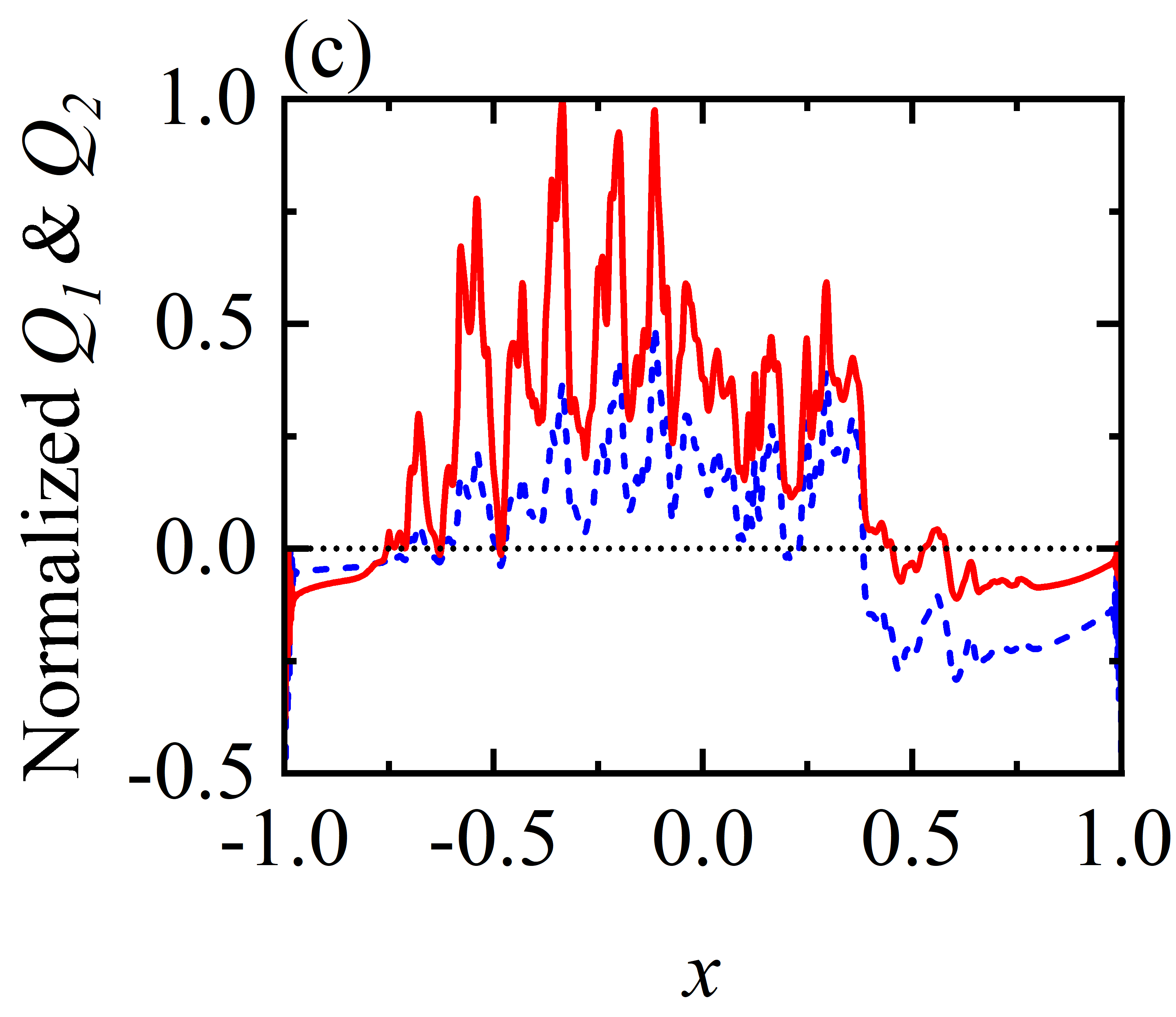}}
\subfigure{
\includegraphics[width=0.32\textwidth]{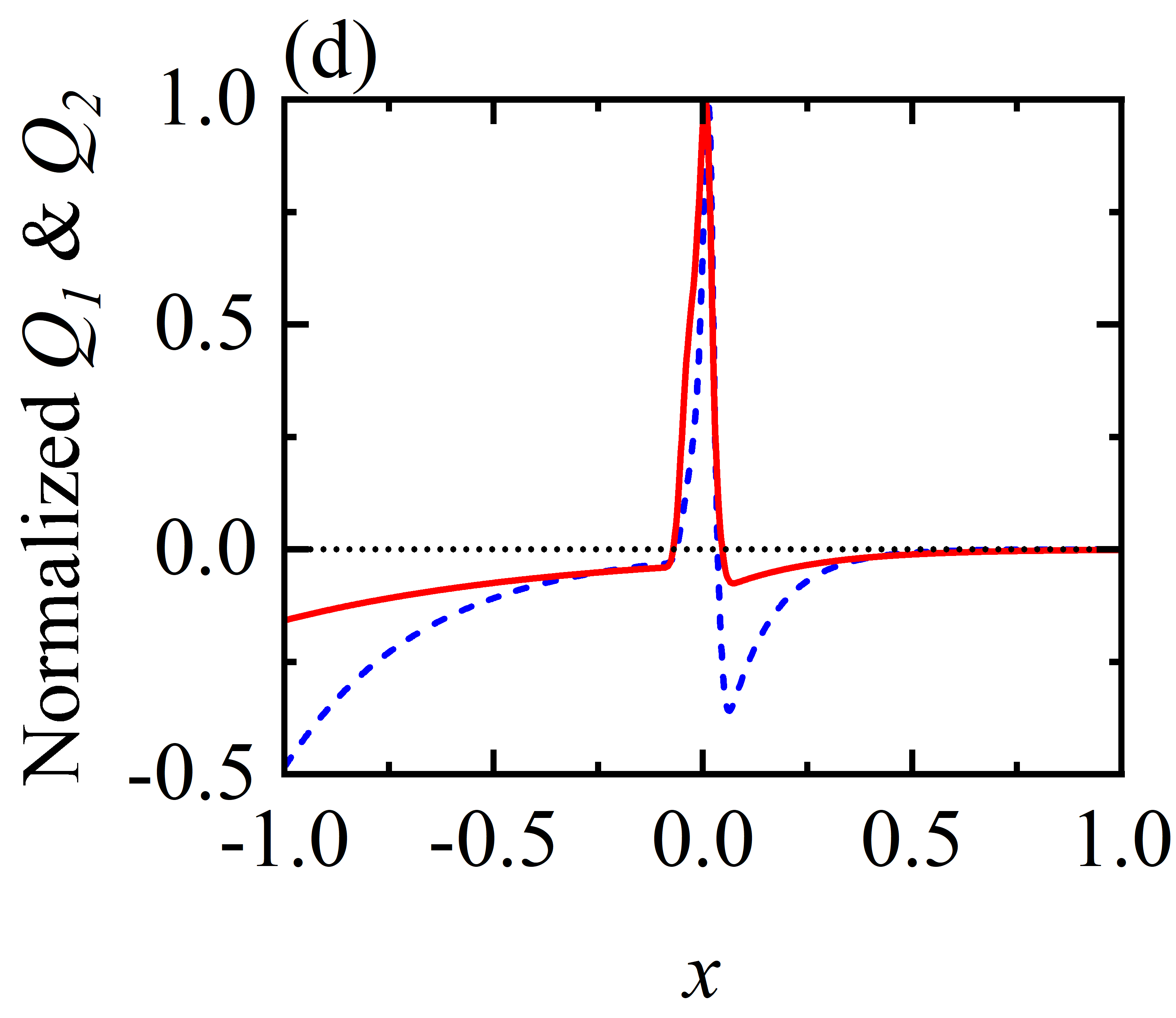}}
\subfigure{
\includegraphics[width=0.32\textwidth]{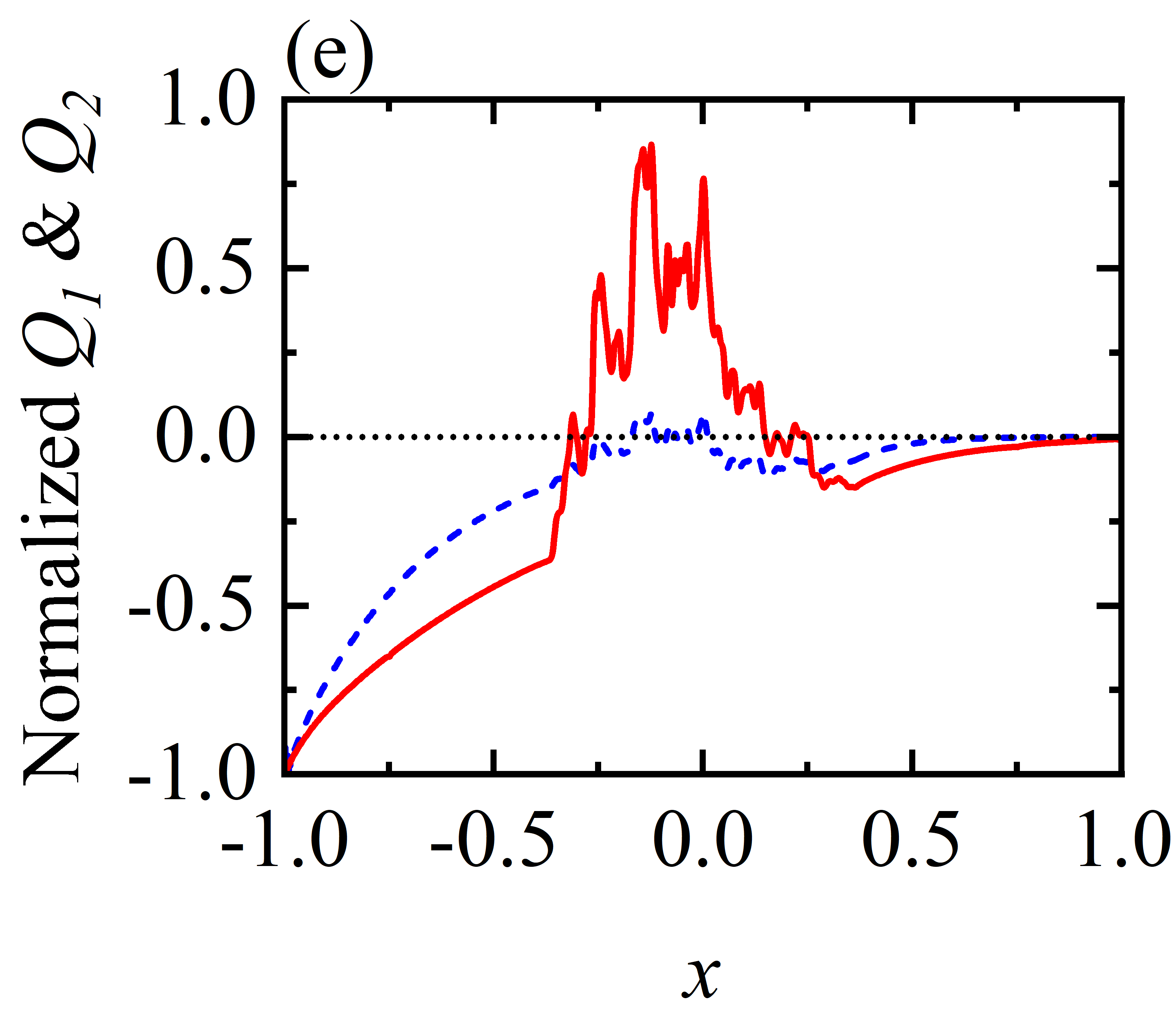}}
\subfigure{
\includegraphics[width=0.32\textwidth]{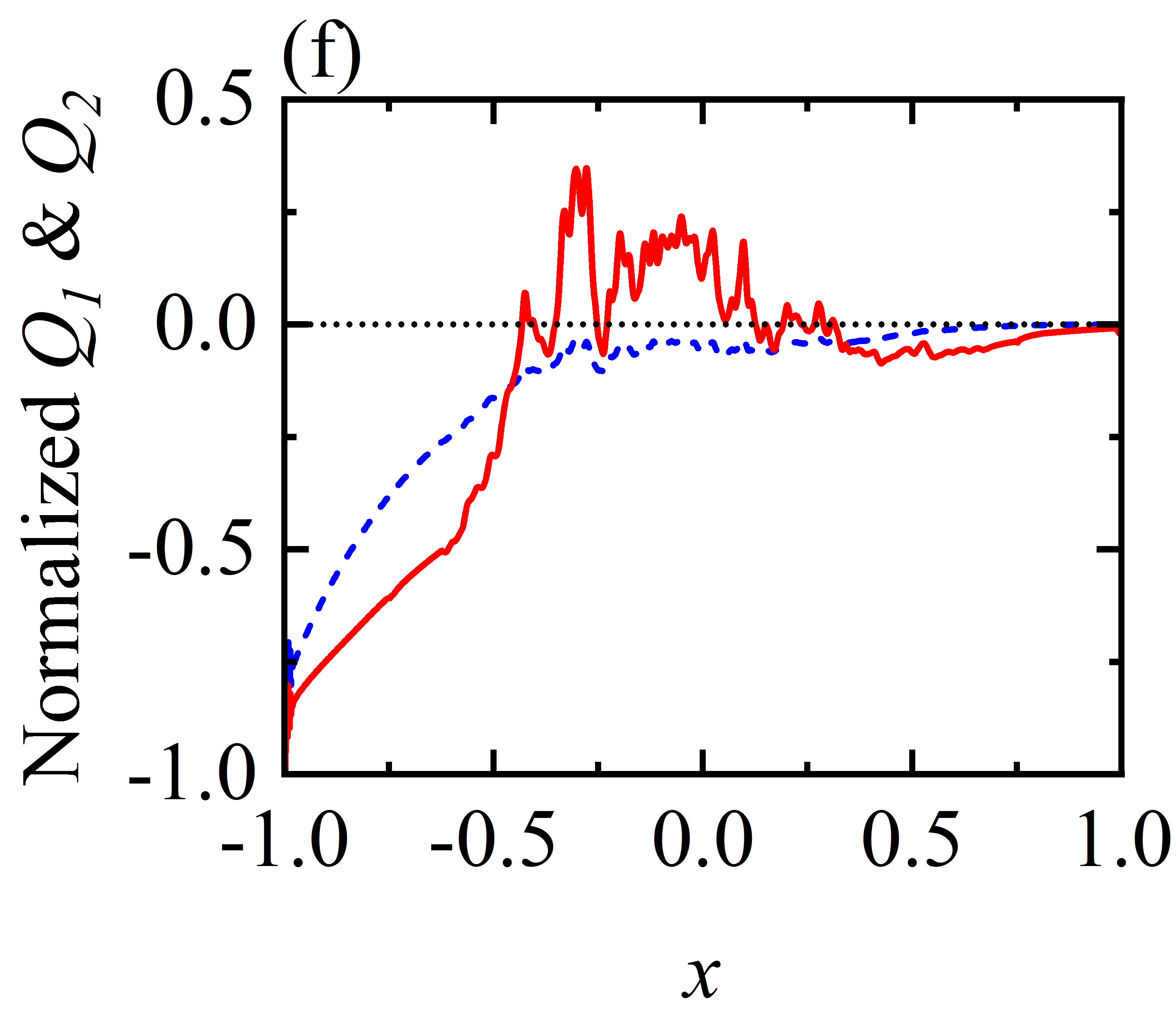}}
\caption{Identification of unstable regions based on the local instability criteria for incompressible flow  $Q_1=-\frac{\partial\bar{\rho}}{\partial x}\frac{\partial\bar{p}}{\partial x}$, and for compressible flow $Q_2=\bar{c}^2\frac{\partial\bar{\rho}}{\partial x}-\frac{\partial\bar{p}}{\partial x}$.
The $Q_2$ is equivalent to the expression (\ref{Q2}) for the RT cases considered here, owing to the negative pressure gradient in gravitational field. 
The ordinate has been normalized by their maximum amplitudes.
Subfigures (a)$\sim$(c) and (d)$\sim$(f) correspond to moments of $t/\tau=$ 1, 2.5 and 4 for the case of $S_r=$ 0.5 and 3 respectively.} 
\label{unstareg}
\end{figure}

We further assess the differences of these criteria in identifying unstable regions by analyzing their spatial distributions along the gravitational direction.
These distributions are obtained through spanwise-plane average of the HiFi  data.
Figure \ref{unstareg} reveals that for the case of $S_r=$ 0.5, both criteria identify identical unstable regions at $t/\tau=$ 1 and 2.5.
However, at  $t/\tau=$ 4, the compressible criterion predicts a significantly wider unstable area than its incompressible counterpart, reflecting the baseline model's under-prediction for late-stage mixing evolution.
For the case of  $S_r=$ 3, both criteria initially agree at $t/\tau=$ 1 due to negligible compressibility effects in early stages. 
However, remarkable differences emerge at $t/\tau=$ 2.5 and 4, where the compressible criterion identifies substantially wider unstable regions. 
These observations thus indicate that the incompressible criterion fails to correctly identify instability when the compressibility effects are significant.
Considering that the instability criterion is essential for RANS models as shown in the closures (\ref{Skf}) and (\ref{Assi}), the preceding analysis suggests that compressibility effects must be incorporated in modeling the turbulent mass flux.

\subsection{Compressibility corrections} \label{Compressibility corrections}
Analysis of the TKE equation budgets from HiFi simulations confirms that the turbulent mass flux  plays a dominant role. 
Since the seminal work of \citet{Dimonte2006K-L}, this term $-\overline{u^{''}_i}\frac{\partial \bar{p}}{\partial x_i}$ has been modeled based on a simple buoyancy-drag model and has received subsequent refinements continuously,  leading to the current form (\ref{Skf}) or  similar variants \citep{Kokkinakis2019Modeling,Kokkinakis2020Two}.  
However, compressibility effect has not been considered yet.
Indeed, the majority of published RANS mixing studies concentrate on incompressible flows, with quite limited attention given to compressible cases.
This section aims to explore how to incorporate compressibility effects into the modeling of the turbulent mass flux term $-\overline{u^{''}_i}\frac{\partial \bar{p}}{\partial x_i}$, with a particular focus on the turbulent mass flux velocity $\overline{u^{''}_i}$, which serves as a critical component.
Utilizing the definitions of Reynolds and Favre averaging, the turbulent mass flux velocity $\overline{u^{''}_i}$ can be expressed as $-\frac{\overline{\rho^{'}u^{'}_{i}}}{\bar{\rho}}$. This expression highlights that $\overline{u^{''}_i}$ fundamentally characterizes the turbulent transport on density fluctuations.
Therefore, it is natural to explore the potential influence factors on density fluctuation in compressible flows.
  
To establish a generalized physical relationship, the  EOS for a perfect gas is employed in the following form for the mixture:
\begin{equation} \label{EOS}
    p=(\Gamma-1)\rho e.
\end{equation}
Average the equation (\ref{EOS}) to give: 
\begin{equation} \label{aaEOS}
    \bar{p}=\overline{(\Gamma-1)\rho e}.
\end{equation}
With the Reynolds decomposition, the instantaneous variables $p$, $\rho$, $\Gamma$, and $e$ can be written as 
$p=\bar{p}+p^{'}$, $\rho=\bar{\rho}+\rho^{'}$,  $\Gamma=\bar{\Gamma}+\Gamma^{'}$, and $e=\bar{e}+e^{'}$, which are substituted into the equations (\ref{EOS}) and (\ref{aaEOS}), then subtracting equation (\ref{aaEOS}) from equation (\ref{EOS}) to yield the linear relation between  fluctuation variables as follows
\begin{equation} \label{rhof}
    \rho^{'}=\frac{\bar{\rho}}{\bar{p}}p^{'}-\frac{\bar{\rho}}{\bar{e}}e^{'}-\frac{\bar{\rho}}{\bar{\Gamma}-1}\Gamma^{'}, 
\end{equation}
where the nonlinear fluctuation correlations are discarded.
The adiabatic index fluctuation $\Gamma^{'}$ can be considered neglected due to the following reasons.
Firstly, for mixtures comprising gases with identical molecular structures,  the adiabatic index $\Gamma_L$ and $\Gamma_H$ of the light and heavy fluids are identical.
Consequently, $\Gamma$ of the mixture is independent of the species and can be treated as a constant,  with $\Gamma^{'}$ becoming appreciable only at extreme temperatures which is beyond the scope of the present study.
Secondly, for mixtures consisting of gases with different molecular structures, the variation in $\Gamma$ of the mixture remains insignificant, even if the adiabatic index $\Gamma_L$ and $\Gamma_H$  of the light and heavy fluids are different and the $\Gamma$ of the mixture  is species-dependent.
This is attributed to the small difference in adiabatic indices for gases with different molecular structures.
For example, $\Gamma=1.4$ for  diatomic gas, $\Gamma=1.33$ for polyatomic gas, it indicates negligible spatial gradients along the direction of mixing development.
In contrast, the fluctuations in density, pressure, and internal energy are substantially larger than  $\Gamma^{'}$, and as thus, the $\Gamma^{'}$ can be considered negligible. 
Consequently, the equation (\ref{rhof}) reduces to:
\begin{equation} \label{rrhof}
    \rho^{'}= \frac{\bar{\rho}}{\bar{p}}p^{'}-\frac{\bar{\rho}}{\bar{e}}e^{'}. 
\end{equation}
The mass flux velocity is then  obtained by multiplying the equation (\ref{rrhof}) by $u^{'}$ and then applying averaging:
\begin{equation}\label{tmf}
     \overline{u_{i}^{''}}=-\frac{\overline{\rho^{'}u^{'}_{i}}}{\bar{\rho}}=-\frac{\overline{p^{'}u^{'}_{i}}}{\bar{p}}+\frac{\overline{e^{'}u^{'}_{i}}}{\bar{e}}.
\end{equation}

\begin{figure} 
\centering
\includegraphics[width=0.75\textwidth]{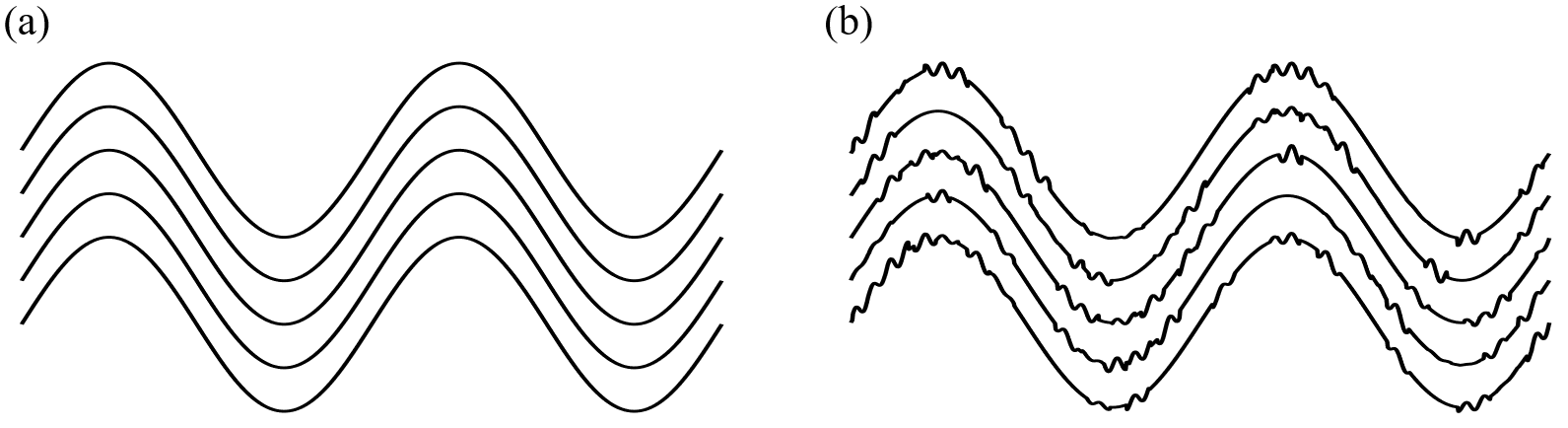}
\caption{Schematic diagrams of two  adjacent equilibrium states. 
The subfigure (a) gives the state $A$, a smooth ensemble average state; the subfigure (b) gives the state $B$, an instantaneous state evolving from the state $A$ under small perturbations.} 
\label{schgibbs}
\end{figure}

The local instability criterion (\ref{Q2}) for compressible flows provides valuable insights for identifying unstable regions, offering critical inspiration for the development of a compressible RANS mixing model. 
This criterion is incorporated into the modeling of the mass flux velocity by combining with the thermodynamic Gibbs relation (\citet{Chassaing2002Variable}, chapter 6):  
\begin{equation} \label{gibbs}
    T\delta s=\delta h-\delta p/\rho,
\end{equation}
where $s$ denotes  entropy.
This relation is applied to the following scenario. 
Consider two adjacent equilibrium states, $A$ and $B$, as illustrated in figure \ref{schgibbs}.
Here, state $A$ represents a smooth ensemble average state, while state $B$ evolves from state $A$ under small perturbations, corresponding to an instantaneous state.
These perturbations are sufficiently small to satisfy the mathematical definition of variation, enabling a continuous transition from state
$A$ to state $B$.
Consequently, the variations  $\delta s$, $\delta h$, and $\delta p$ can be equated to the corresponding small perturbations $s^{'}$, $h^{'}$ and $p^{'}$ respectively, yielding:
\begin{equation} \label{flugibbs}
    Ts^{'}=h{'}-p{'}/\rho.
\end{equation}
Substitute $T=\bar{T}+T^{'}$ and $
\rho=\bar{\rho}+\rho^{'}$ into the equation (\ref{flugibbs}) to yield the linear relation between  fluctuation variables as follows
\begin{equation} \label{agibbs}
    \bar{T}s^{'}=h^{'}-p^{'}/\bar{\rho},
\end{equation}
where the nonlinear fluctuation correlations are discarded.

Next, substituting  the pressure fluctuation in equation (\ref{tmf}) with (\ref{agibbs}) yields:
\begin{equation}\label{tost}
    \overline{u_{i}^{''}}=-\frac{\overline{\rho^{'}u^{'}_{i}}}{\bar{\rho}}=-\frac{\bar{\rho}}{\bar{p}}\overline{h^{'}{u^{'}_{i}}-\bar{T}s^{'}u^{'}_{i}}+\frac{\overline{e^{'}u^{'}_{i}}}{\bar{e}}\approx\frac{\bar{\rho}\bar{T}}{\bar{p}}\overline{s^{'}u^{'}_{i}}-\frac{\bar{\rho}}{\bar{p}}\overline{e^{'}u^{'}_{i}},
\end{equation}
where $h=\Gamma e$ is used, and the approximation arises from neglecting a third-order fluctuation correlation term and a $\Gamma^{'}$ term.
These two terms in (\ref{tost}) will be modeled respectively.

Using the relation $s=c_vln\frac{p}{\rho^{\Gamma}}$,  $\frac{\bar{\rho}\bar{T}}{\bar{p}}\overline{s^{'}u^{'}_{i}}$ can be expressed as 
\begin{equation} \label{Sf}
     \frac{\bar{\rho}\bar{T}}{\bar{p}}\overline{s^{'}u^{'}_{i}}\approx\frac{\bar{\rho}\bar{T}\bar{c}_{v}}{\bar{p}}\overline{\left(ln\frac{p}{\rho^{\Gamma}}\right)^{'}u^{'}_{i}},
\end{equation}
where the fluctuations involving $c_v^{'}$ are ignored, similar to the treatment of \citet{Franklin2010Simulation}. 
The expression (\ref{Sf}) can be written as 
\begin{equation}\label{sres}
   \frac{\bar{\rho}\bar{T}\bar{c}_{v}}{\bar{p}}\overline{\left(ln\frac{p}{\rho^{\Gamma}}\right)^{'}u^{'}_{i}}=\frac{\bar{\rho}\bar{T}\bar{c}_{v}}{\bar{p}}\overline{(lnp-\Gamma ln\rho)^{'}u^{'}_{i}}\approx\frac{\bar{\rho}\bar{T}\bar{c}_{v}}{\bar{p}}\overline{(lnp)^{'}u^{'}_{i}-\bar{\Gamma}(ln\rho)^{'}u^{'}_{i}},     
\end{equation} 
Where the term involving $\Gamma^{'}$ is ignored, as discussed  earlier.
The $(lnp)^{'}$ and $(ln\rho)^{'}$ are the fluctuation parts of $lnp$ and $ln\rho$ respectively, and can be obtained based on the following operation.
Implement  Taylor expansions for  $lnp$ and $ln\rho$ surrounding with the mean parts $\bar{p}$ and $\bar{\rho}$, reading $lnp=ln(\bar{p}+p^{'}) \approx ln\bar{p}+p^{'}/\bar{p}+O({p^{'}}^{2})$, and $ln\rho=ln(\bar{\rho}+\rho^{'}) \approx ln\bar{\rho}+\rho^{'}/\bar{\rho}+O({\rho^{'}}^{2})$, thus yielding $(lnp)^{'}\approx p^{'}/\bar{p}$ and $(ln\rho)^{'}\approx\rho^{'}/\bar{\rho}$ by discarding the nonlinear fluctuation correlations.
Consequently, the expression (\ref{sres}) is transformed as
\begin{equation}\label{fisf}
  \frac{\bar{\rho}\bar{T}\bar{c}_{v}}{\bar{p}}\overline{(lnp)^{'}u^{'}_{i}-\bar{\Gamma}(ln\rho)^{'}u^{'}_{i}}=\frac{\bar{\rho}\bar{T}\bar{c}_{v}}{\bar{p}}\overline{p^{'}u^{'}_{i}/\bar{p}-\bar{\Gamma}\rho^{'}u^{'}_{i}/\bar{\rho}}.  
\end{equation}
Based on the GDA, the (\ref{fisf}) can be closed with the mean variables, i.e.  
\begin{equation}\label{sclo}
   \frac{\bar{\rho}\bar{T}\bar{c}_{v}}{\bar{p}}\overline{p^{'}u^{'}_{i}/\bar{p}-\bar{\Gamma}\rho^{'}u^{'}_{i}/\bar{\rho}}={{C}_{3}}{{\nu }_{t}} 
    \frac{\bar{\rho}\bar{T}\bar{c}_{v}}{\bar{p}}\left (-\frac{1}{\bar{p}}\frac{\partial \bar{p}}{\partial x_{i}}+\frac{\bar{\Gamma
    }}{\bar{\rho}}\frac{\partial\bar{\rho}}{\partial x_{i}}\right)=
    {{C}_{3}}{{\nu }_{t}}\frac{{\bar{\Gamma }}}{\bar{\rho }\left( \bar{\Gamma }-1 \right)}\left( \frac{\partial \bar{\rho }}{\partial x_{i}}-\frac{1}{{{{\bar{c}}}^{2}}}\frac{\partial \bar{p}}{\partial x_{i}} \right),     
\end{equation} 
where $C_3$ is a model coefficient to be determined.
It is evident that the closure in (\ref{sclo}) is intrinsically linked to the compressibility criterion (\ref{Q2}), highlighting the physical consistency of the proposed formulation.

Using the relation $e=c_vT$, $-\frac{\bar{\rho}}{\bar{p}}\overline{e^{'}u^{'}_{i}}$ can be expressed as 
\begin{equation} \label{Tf}
    -\frac{\bar{\rho}}{\bar{p}}\overline{e^{'}u^{'}_{i}}\approx-\frac{\bar{\rho}\bar{c}_{v}}{\bar{p}}\overline{T^{'}u^{'}_{i}},
\end{equation}
where  $c_v^{'}$ is ignored, similar to the treatment of (\ref{Sf}). 
The pure turbulent heat flux $\overline{T^{'}u^{'}_{i}}$ in  (\ref{Tf}) is straightforward modeled based on the expression proposed by \citet{osti2003Compressibility}, reading
\begin{equation}\label{thfm}
    \overline{T^{'}u^{'}_{i}}=-C_4v_{t}\left(\frac{\partial \bar{T}}{\partial x_{i}}-\frac{\bar{\Gamma}-1}{\bar{\Gamma}}\frac{\bar{T}}{\bar{p}}\frac{\partial \bar{p}}{\partial x_{i}}\right),
\end{equation}
where  $C_4$ is a model coefficients to be determined.
By combining with the (\ref{thfm}), the $-\frac{\bar{\rho}\bar{c}_{v}}{\bar{p}}\overline{T^{'}u^{'}_{i}}$  is expressed as
\begin{equation}\label{Tclo}
    -\frac{\bar{\rho}\bar{c}_{v}}{\bar{p}}\overline{T^{'}u^{'}_{i}}={{C}_{4}}{{\nu }_{t}}\frac{{{{\bar{c}}}_{p}}}{\bar{\rho }{{{\bar{c}}}^{2}}}\left( \bar{\rho }\frac{\partial \bar{T}}{\partial x_{i}}-\frac{1}{{{{\bar{c}}}_{p}}}\frac{\partial \bar{p}}{\partial x_{i}} \right).
\end{equation}
It is worth emphasizing that this formulation of (\ref{thfm}) deviates from the conventional GDA framework by explicitly accounting for counter-gradient heat transport mechanisms. 
Classical gradient diffusion theories, such as Fourier's law, posit that heat transfer occurs from high-temperature to low-temperature regions, with flux magnitude proportional to the local temperature gradient.
However, turbulent flow dynamics can induce anomalous  heat transport phenomena where heat is transferred against the temperature gradient---from low-temperature to high-temperature regions.
Such a counter-gradient behavior is ubiquitously observed in convection-dominated boundary layer, buoyancy-dominated turbulence, turbulent coherent structures, among other scenarios.

\begin{figure} 
\centering
\subfigure{
\includegraphics[width=0.45\textwidth]{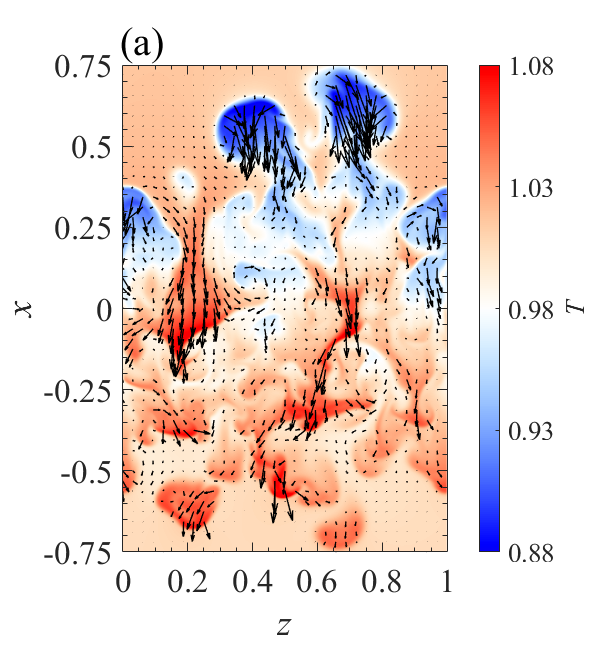}}
\hspace{0.2cm}
\subfigure{
\includegraphics[width=0.45\textwidth]{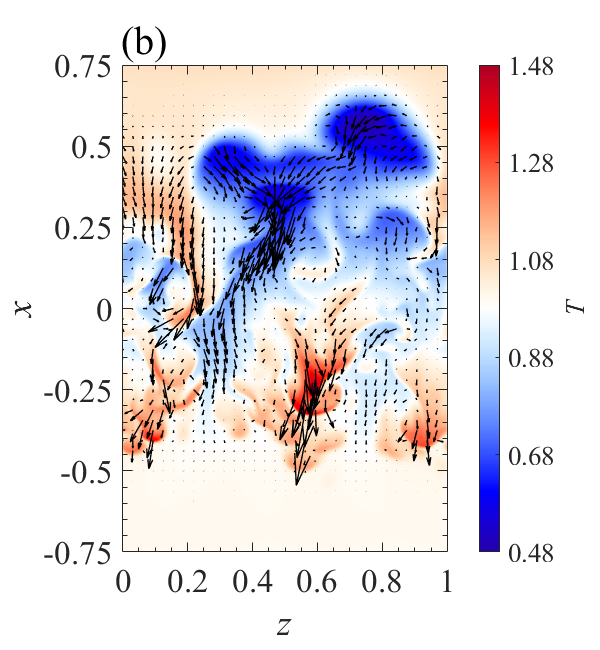}}
\caption{Contours of instantaneous temperature in the central $x-z$ plane  for cases of (a) $S_r=0.5$  and (b) 3  at $t/\tau=4$.
The instantaneous turbulent heat flux is also displayed with the vector arrows.} 
\label{contours}
\end{figure}

In the density-stratified compressible RT mixing flows investigated in this study, counter-gradient heat transfer is observed, as  clearly evidenced by the flow field visualization in figure \ref{contours}.
To elucidate this phenomenon, we analyze the late-stage flow field at $t/\tau=4$ from the central $x-z$ plane of the 3D HiFi simulations for weak ($S_r=0.5$) and strong ($S_r=3$) stratification cases.
The visualization employs a background temperature color map overlaid with instantaneous turbulent heat flux vectors $\overline{T^{'}u^{'}_{i}}$.
Figure \ref{contours} reveals the following points.
Firstly, the heavy fluid regions generally exhibit lower temperatures compared to the light fluid regions, and moreover, density stratification promotes a wider range of temperature variations across the mixing zone. 
Secondly, strong density stratification ($S_r=3$) contributes to produce organized large-scale flow structures, enhancing heat transfer efficiency.
The case with weak density stratification ($S_r=0.5$) displays dispersed small-scale turbulent structures, resulting in heat transfer dominated by local mixing.
Last but not least, the turbulent heat flux shows significant deviation from the local temperature gradient direction, with clear evidence of heat transfer from low-temperature to high-temperature regions.
These findings  underscore the importance of turbulent heat flux in the flows considered here, and further necessitate explicit consideration of counter-gradient diffusion in modeling $\overline{T^{'}u^{'}_{i}}$.
Consequently,  based on the work of \citet{osti2003Compressibility}, we adopt the closure formulation (\ref{thfm}) to  address the specific requirements of modeling contour-gradient diffusion in density-stratified compressible RT flows.

Finally, collecting the expressions (\ref{Tclo}) and (\ref{sclo}), the turbulent mass flux velocity is modeled as  
\begin{equation}\label{puf1}
    -\frac{\overline{\rho^{'}u^{'}_{i}}}{\bar{\rho}}={{C}_{3}}{{\nu }_{t}}\frac{{\bar{\Gamma }}}{\bar{\rho }\left( \bar{\Gamma }-1 \right)}\left( \frac{\partial \bar{\rho }}{\partial x_{i}}-\frac{1}{{{{\bar{c}}}^{2}}}\frac{\partial \bar{p}}{\partial x_{i}} \right)+{{C}_{4}}{{\nu }_{t}}\frac{{{{\bar{c}}}_{p}}}{\bar{\rho }{{{\bar{c}}}^{2}}}\left( \bar{\rho }\frac{\partial \bar{T}}{\partial x_{i}}-\frac{1}{{{{\bar{c}}}_{p}}}\frac{\partial \bar{p}}{\partial x_{i}} \right).
\end{equation}
Based on a principle of minimal modifications to the baseline model, the  expression (\ref{puf1}) is integrated into  the baseline closure (\ref{Assi}) for the local Atwood number $A_{ssi}$, yielding:
\begin{equation} \label{newAssi}
   {{A}_{ssi}}=\frac{{\tilde{L}}}{\bar{\rho }+\tilde{L}\left| \frac{\partial \bar{\rho }}{\partial x_{i}} \right|}\left\{ {{C}_{3}} \left[ \frac{{\bar{\Gamma }}}{\bar{\Gamma }-1}\left( \frac{\partial \bar{\rho }}{\partial x_{i}}-\frac{1}{{{{\bar{c}^{2}}}}}\frac{\partial \bar{p}}{\partial x_{i}} \right) \right]+{{C}_{4}}\frac{1}{{{{\bar{c}^{2}}}}} \left( \bar{\rho }{{{\bar{c}}}_{p}}\frac{\partial \bar{T}}{\partial x_{i}}-\frac{\partial \bar{p}}{\partial x_{i}} \right) \right\}. 
\end{equation} 
To ensure consistency with the baseline form (\ref{Assi}) in the incompressible limit $\bar{c}\rightarrow \infty$, the model coefficient ${{C}_{3}}$ is required to meet the constraint
\begin{equation}
    C_3=\frac{C_{A}(\bar{\Gamma }-1)}{{\bar{\Gamma }}}. 
\end{equation}
Additionally, it is found that the present model can achieve optimal predictions when $C_4=C_3$, based on calibrations with HiFi data.
Consequently, $C_4=C_3=\frac{C_{A}(\bar{\Gamma }-1)}{{\bar{\Gamma }}}$ is adopted in this study.

Finally, the local Atwood number $A_{ssi}$ is expressed as
\begin{equation}\label{finaAssi}
    {{A}_{ssi}}=\frac{{\tilde{L}}}{\bar{\rho }+\tilde{L}\left| \frac{\partial \bar{\rho }}{\partial x} \right|}
    \left[ C_A\frac{\partial \bar{\rho }}{\partial x_{i}}+\underbrace{\frac{C_{A}(\bar{\Gamma }-1)}{{\bar{\Gamma }}}\frac{1}{{{{\bar{c}^{2}}}}}\left( \bar{\rho }{{{\bar{c}}}_{p}}\frac{\partial \bar{T}}{\partial x_{i}}-\frac{2\bar{\Gamma }-1}{\bar{\Gamma }-1}\frac{\partial \bar{p}}{\partial x_{i}} \right)}_{compressibility \quad corrections} \right].
\end{equation}
The second term in the expression (\ref{finaAssi}) incorporates all compressibility corrections proposed here. 
Remarkably, this formulation maintains the original model coefficient $C_A$ without introducing additional parameters, thereby preserving the simplicity of this model. 
In the incompressible limit $c\rightarrow0$, the expression (\ref{finaAssi}) naturally converges to its incompressible counterpart given by the formula (\ref{Assi}), demonstrating consistency with the baseline formulation.

The proposed modeling strategy reveals that influences of the turbulent mass flux on evolution of compressible mixing flows predominantly stem from two key factors:  turbulent entropy flux and  turbulent heat flux. 
The present scheme physically captures density variations induced by both entropy and temperature fluctuations.
From a perspective of RANS modeling, the proposed closure takes into accounts for two essential aspects of compressible mixing flows.
Firstly,  the conventional density gradient in the baseline model is replaced by the more physically representative entropy gradient.
Secondly, turbulent heat effects are explicitly incorporated through a modified temperature gradient diffusion formulation that accounts for counter-gradient diffusion effect.
These modeling considerations are expected to provide a comprehensive framework for capturing compressibility effects in variable-density compressible turbulent mixing flows.

\section{Model validation} \label{Model validation}
To evaluate the proposed compressibility corrections, density-stratified compressible RT mixing flows with different stratification parameters $S_r$ are tested. 
The flow configurations, including model coefficients and initialization of turbulence model variables, remain consistent with the baseline model, as stated in section \ref{Convective instability in density-stratified compressible RT flow}.

\begin{figure} 
\centering
\subfigure{
\includegraphics[width=0.45\textwidth,height=0.4\textwidth]{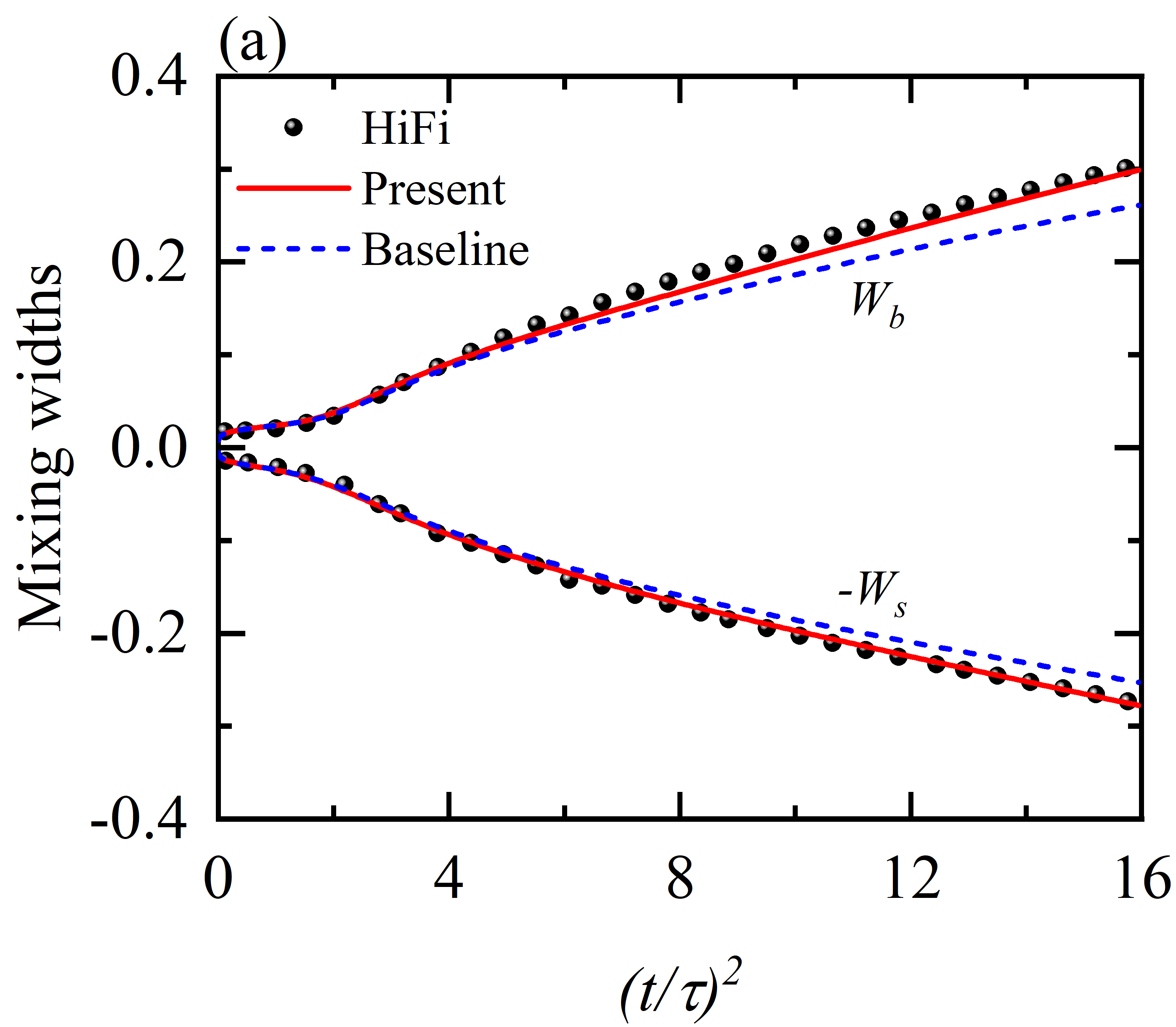}}
\subfigure{
\includegraphics[width=0.45\textwidth,height=0.4\textwidth]{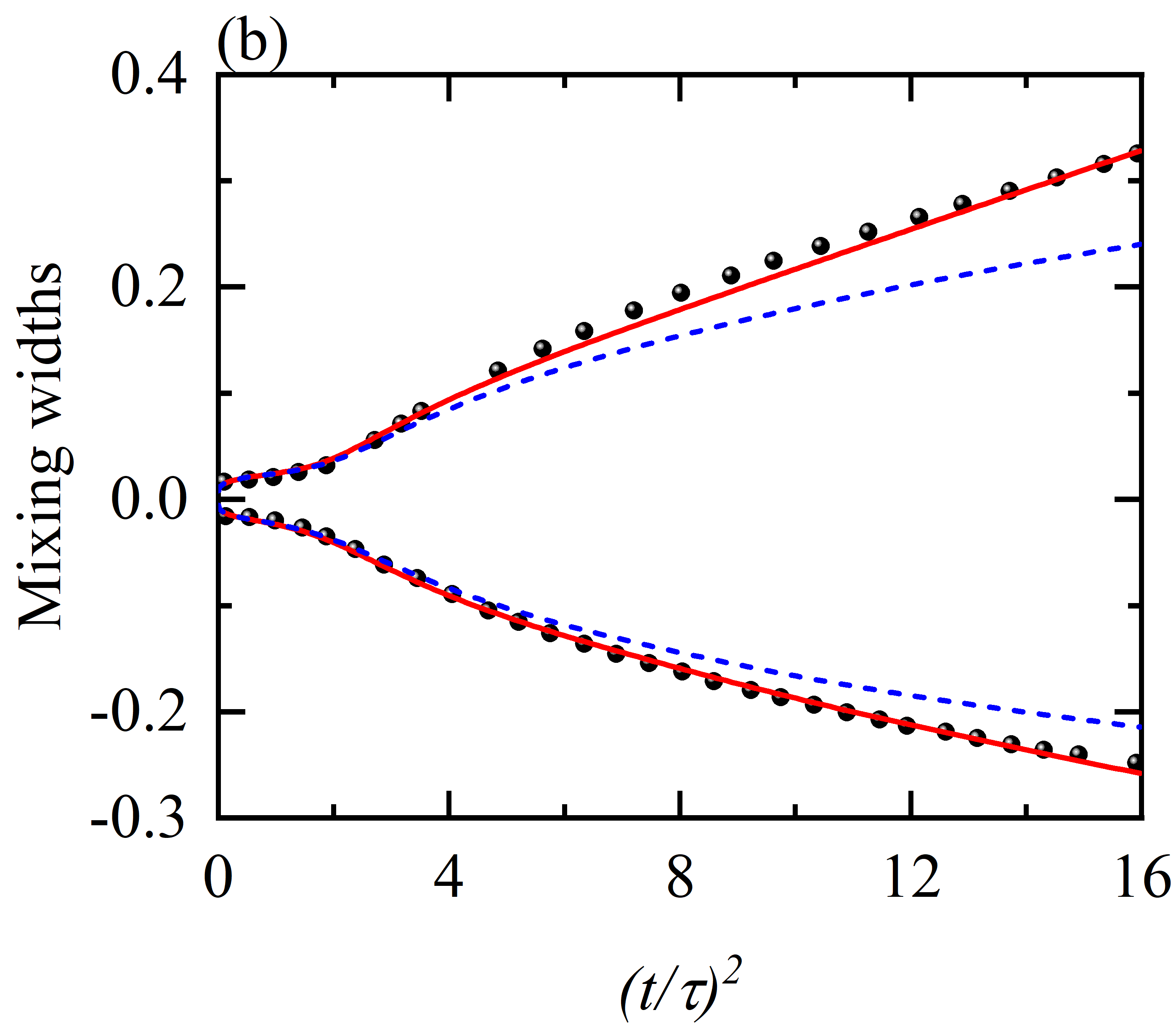}} 
\subfigure{
\includegraphics[width=0.45\textwidth,height=0.4\textwidth]{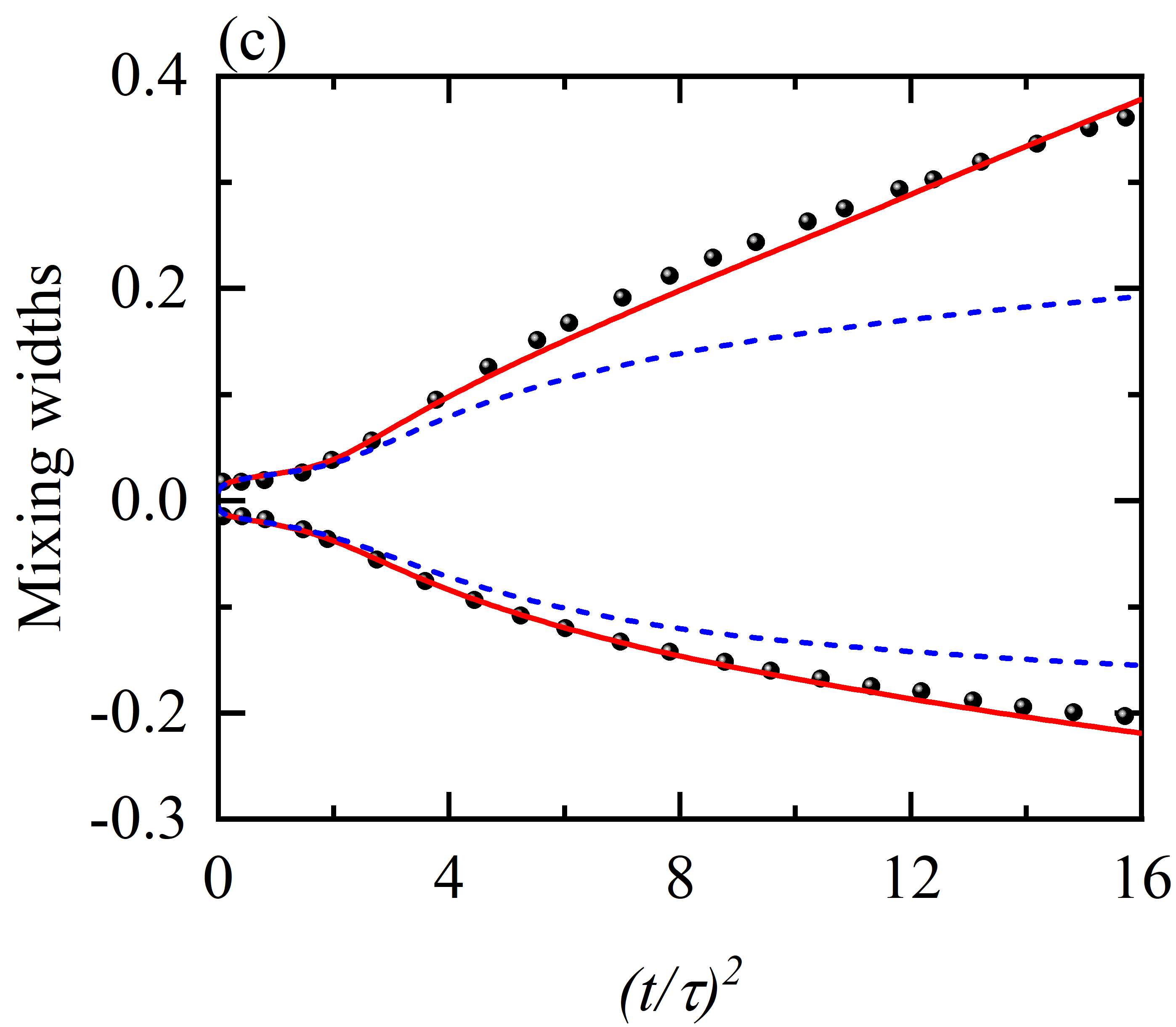}}
\subfigure{
\includegraphics[width=0.45\textwidth,height=0.4\textwidth]{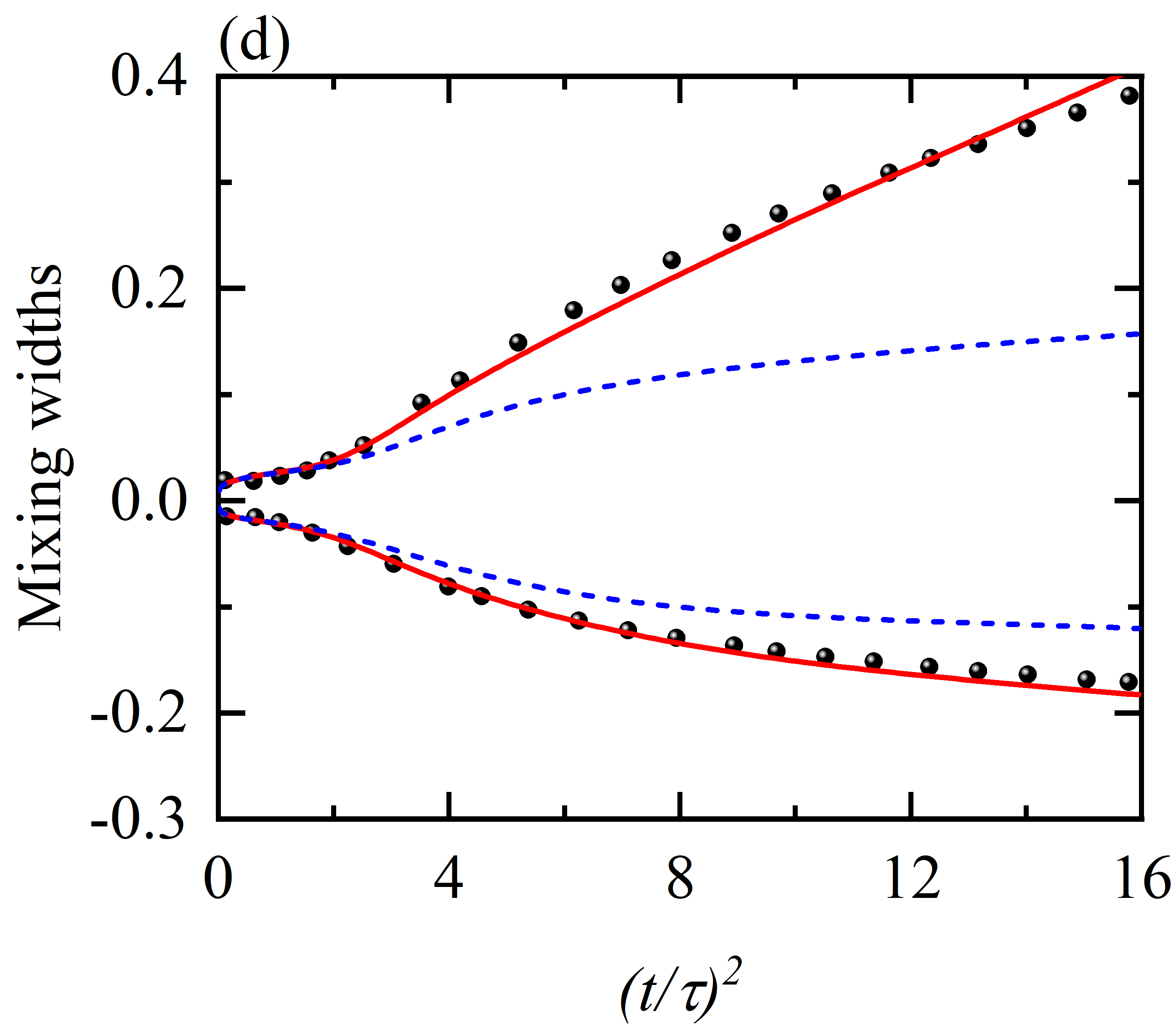}}
\caption{Temporal evolutions of the  mixing widths.
Subfigures (a) to (d) correspond to the cases of $S_r=$ 0.5, 1, 2, and 3, respectively.} 
\label{w}
\end{figure}

Figure \ref{w} illustrates  the temporal evolutions of  bubble ($W_b$) and spike ($W_s$) mixing widths for 4 cases with density stratification parameters $S_r$ ranging from 0.5 to 3.
The proposed compressibility modifications show remarkable consistency with HiFi simulation results across all tested cases.
This improved performance stems from two fundamental  physical considerations: incorporation of the local instability criterion for compressible flows, and  appropriate treatment of turbulent heat flux contributions.
By incorporating these effects, the model achieves a more comprehensive representation of the turbulence production term in the TKE equation, leading to significantly improved predictions of compressible mixing evolution.
It should be noted that identical initialization of the turbulent model variables and model coefficients are maintained throughout all the cases.
The model's robustness is particularly evident in strongly stratified cases ($S_r$=3), where it still delivers accurate predictions despite the significant compressibility effect.
The success of the compressible closure (\ref{finaAssi}) without additional empirical coefficients stems from its physically grounded formulation, which effectively captures the essential dynamics of compressible mixing flows.

\begin{figure} 
\centering
\subfigure{
\includegraphics[width=0.45\textwidth,height=0.4\textwidth]{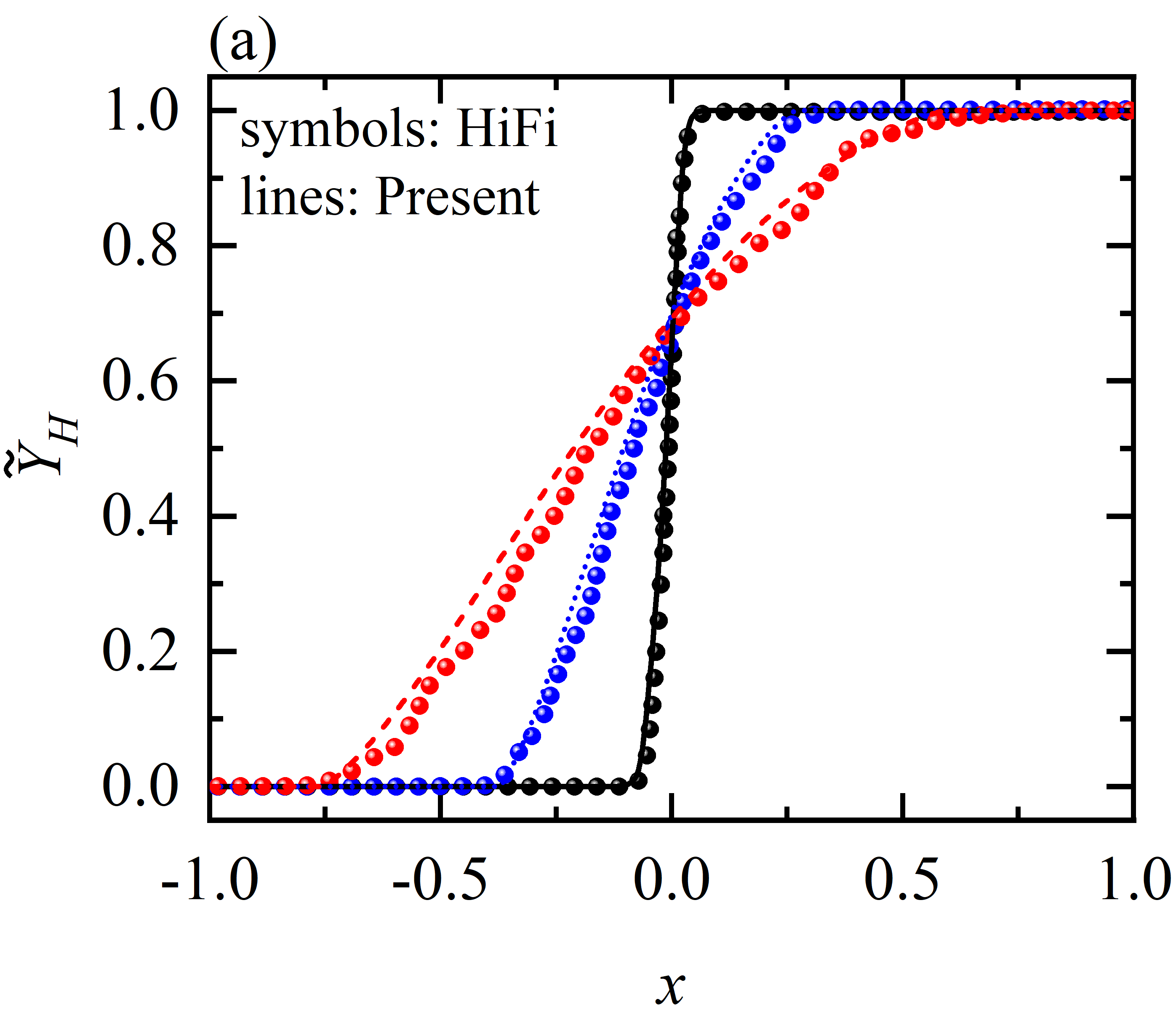}}
\subfigure{
\includegraphics[width=0.45\textwidth,height=0.4\textwidth]{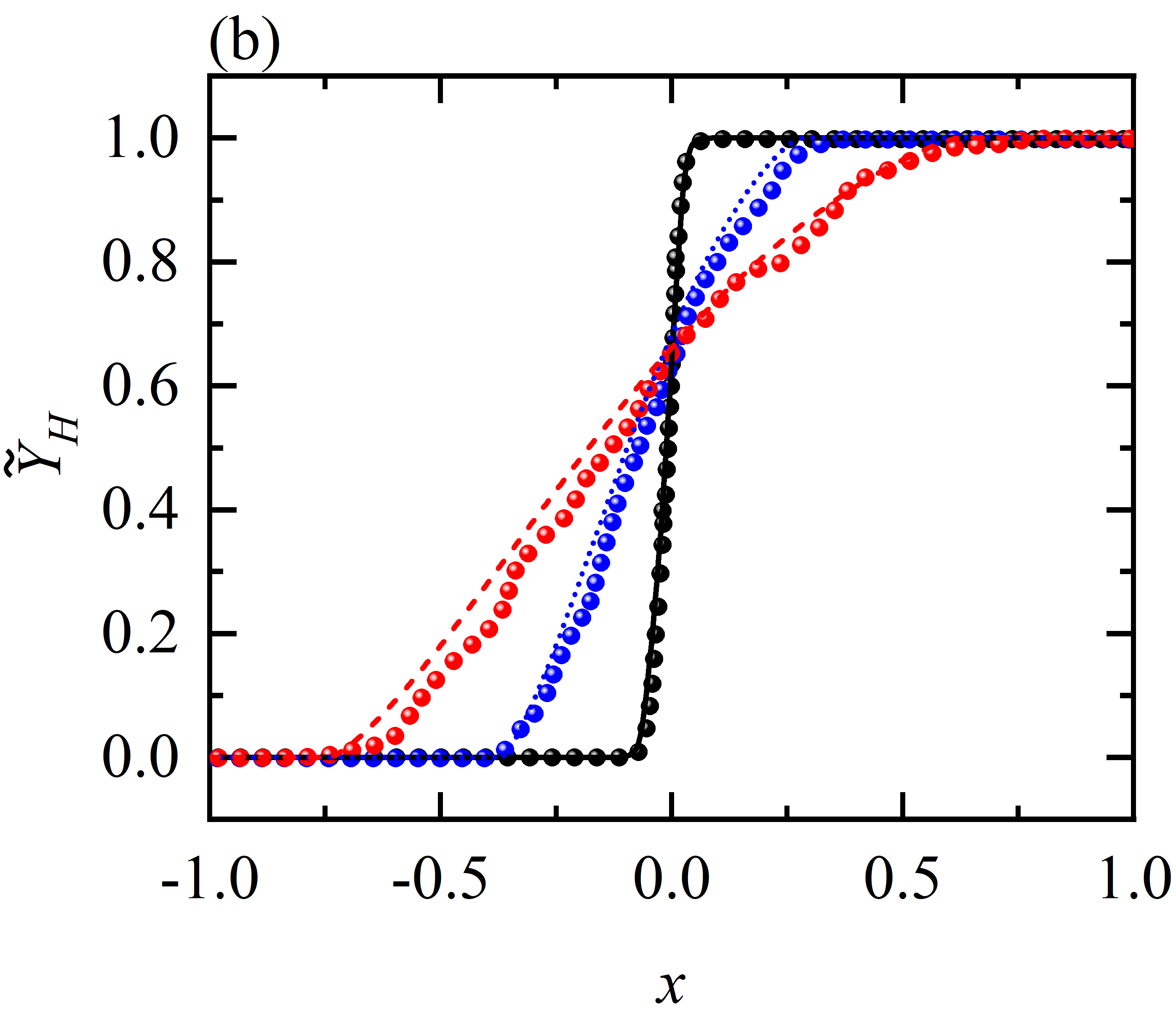}} 
\subfigure{
\includegraphics[width=0.45\textwidth,height=0.4\textwidth]{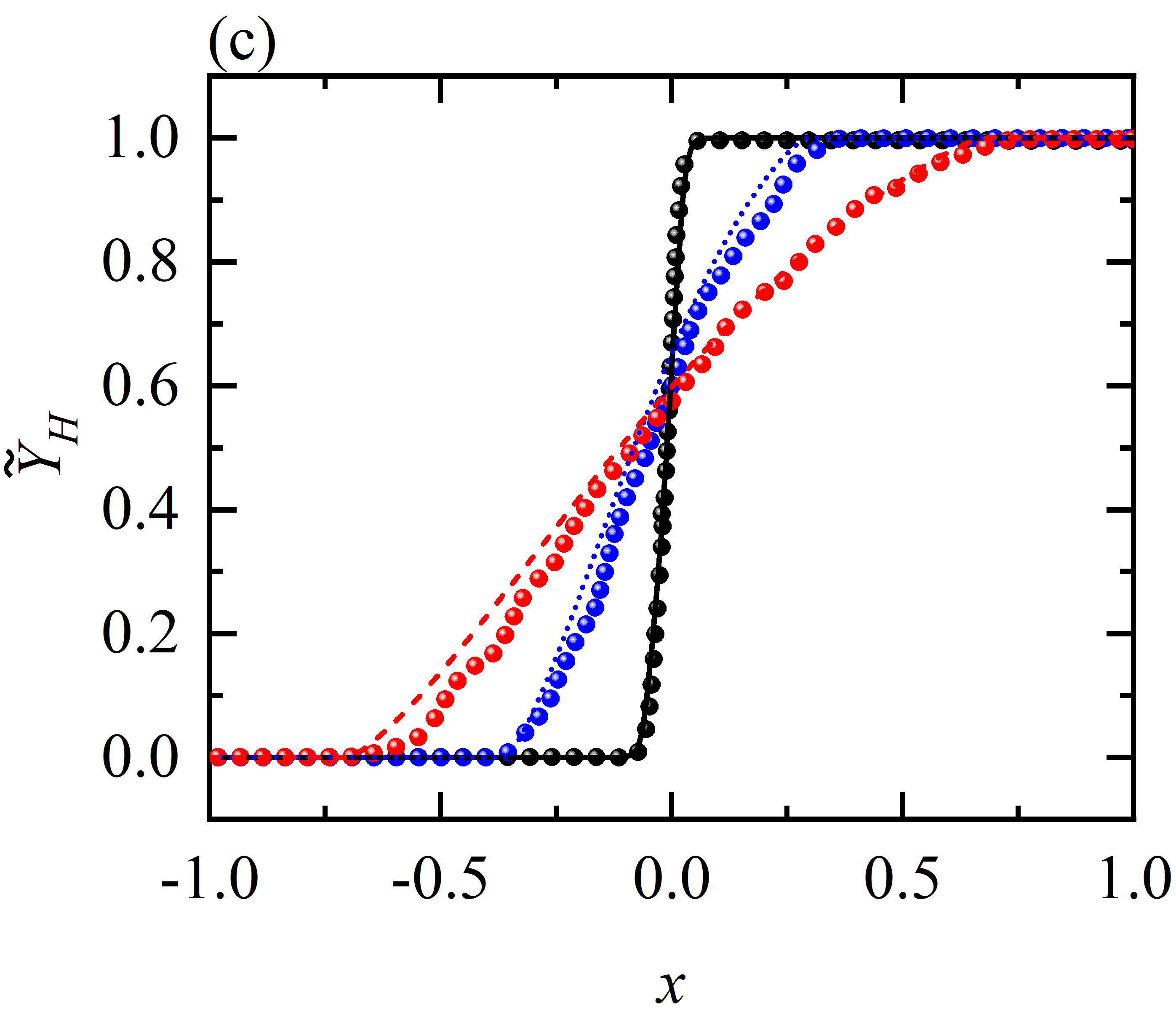}}
\subfigure{
\includegraphics[width=0.45\textwidth,height=0.4\textwidth]{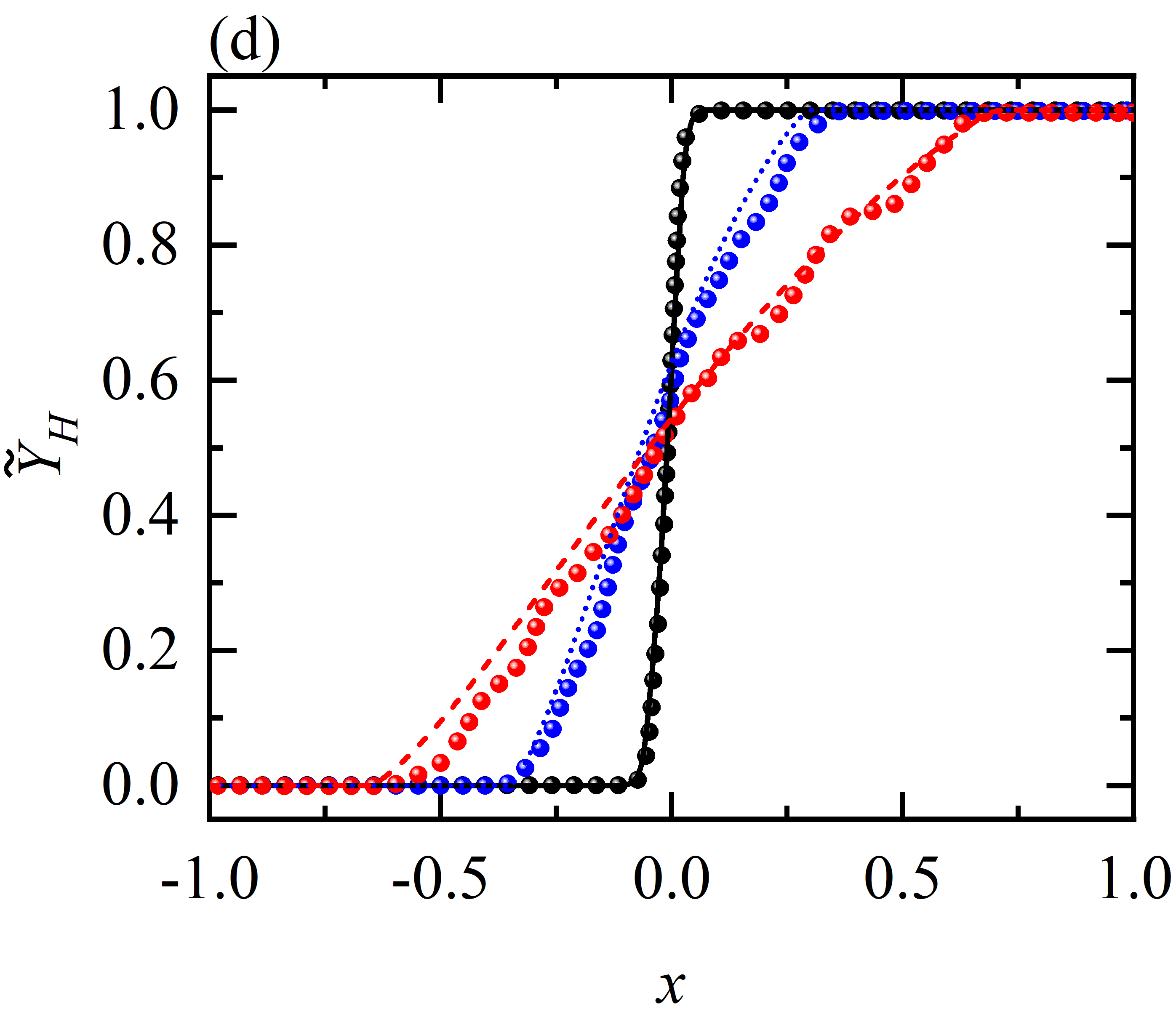}}
\caption{Spatial profiles of the mean species at three different moments $t/\tau=1$ (colored by black), $t/\tau=2.5$ (colored by blue), and $t/\tau=4$ (colored by red). Subfigures (a) to (d) correspond to the cases with $S_r=$ 0.5, 1, 2, and 3, respectively.  } 
\label{speciespro}
\end{figure}

\begin{figure} 
\centering
\subfigure{
\includegraphics[width=0.45\textwidth,height=0.4\textwidth]{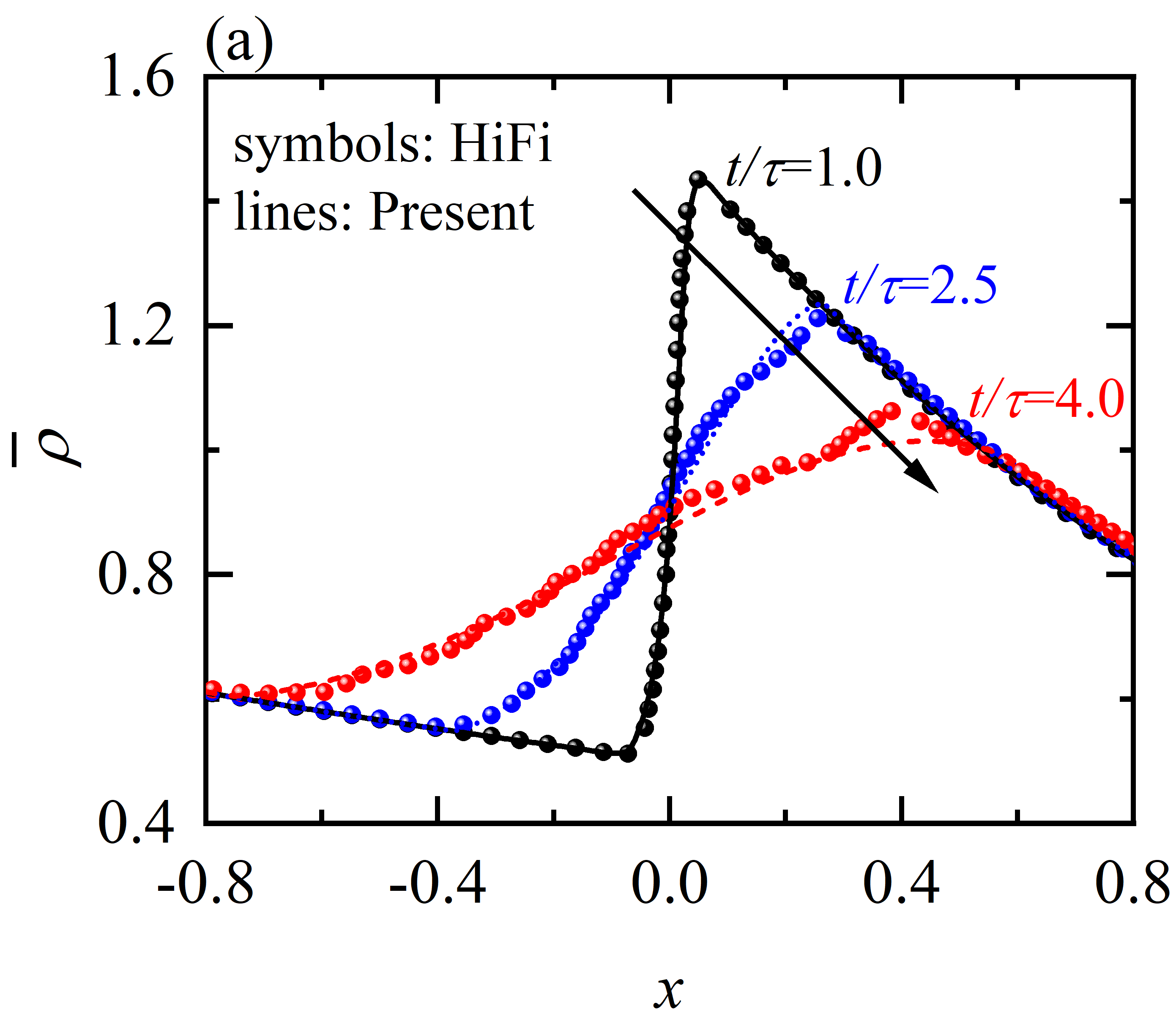}}
\subfigure{
\includegraphics[width=0.45\textwidth,height=0.4\textwidth]{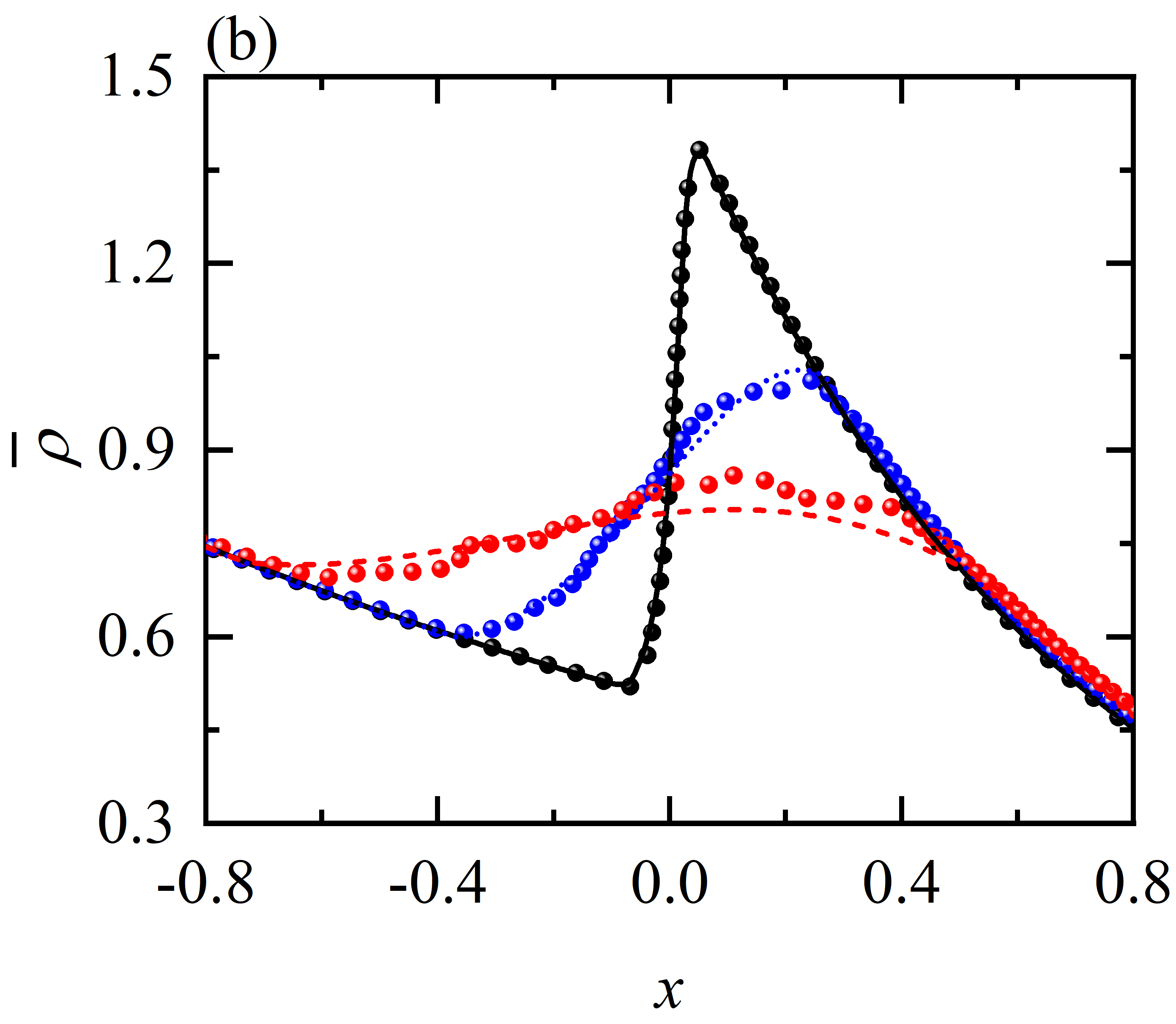}} 
\subfigure{
\includegraphics[width=0.45\textwidth,height=0.4\textwidth]{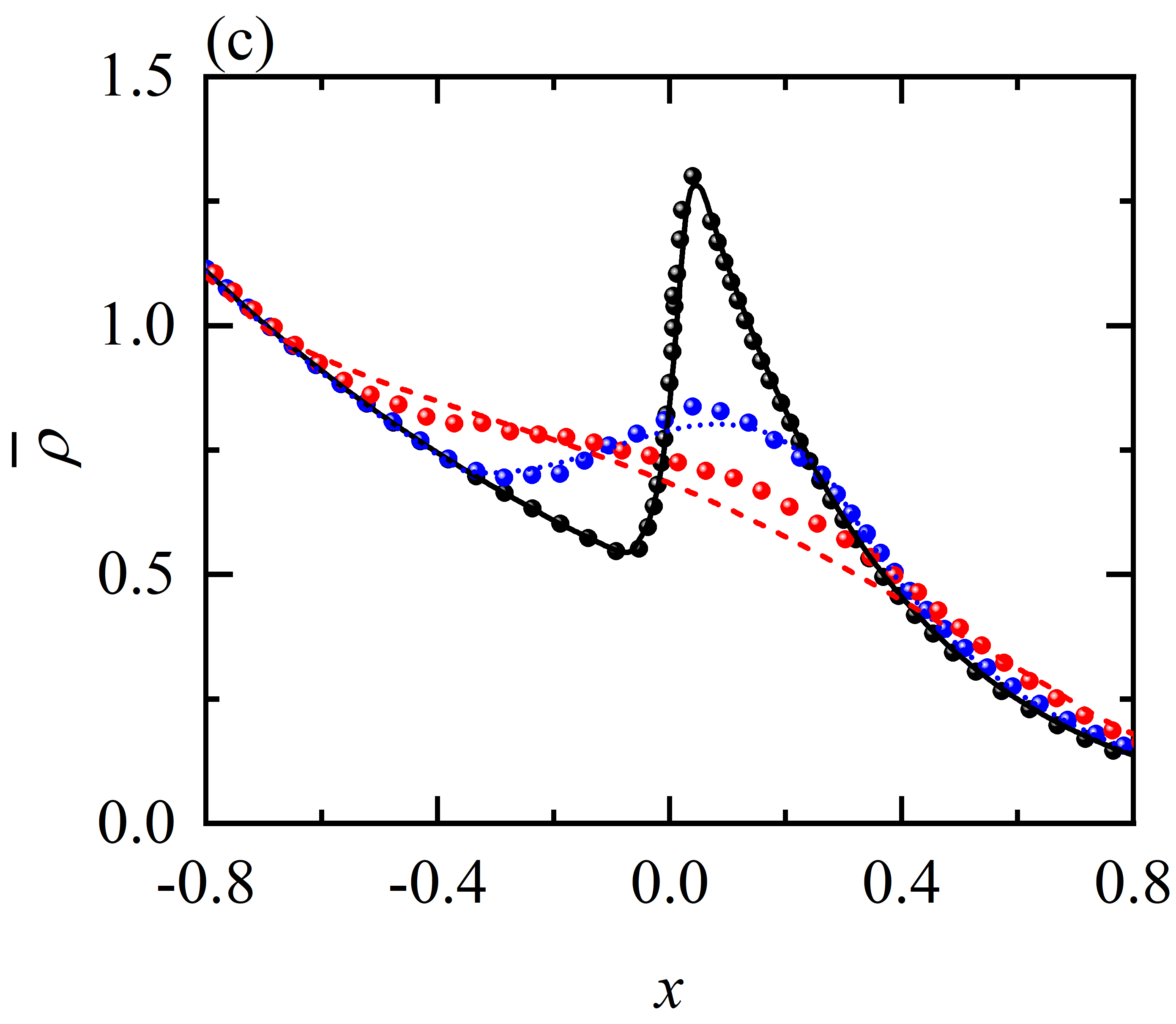}}
\subfigure{
\includegraphics[width=0.45\textwidth,height=0.4\textwidth]{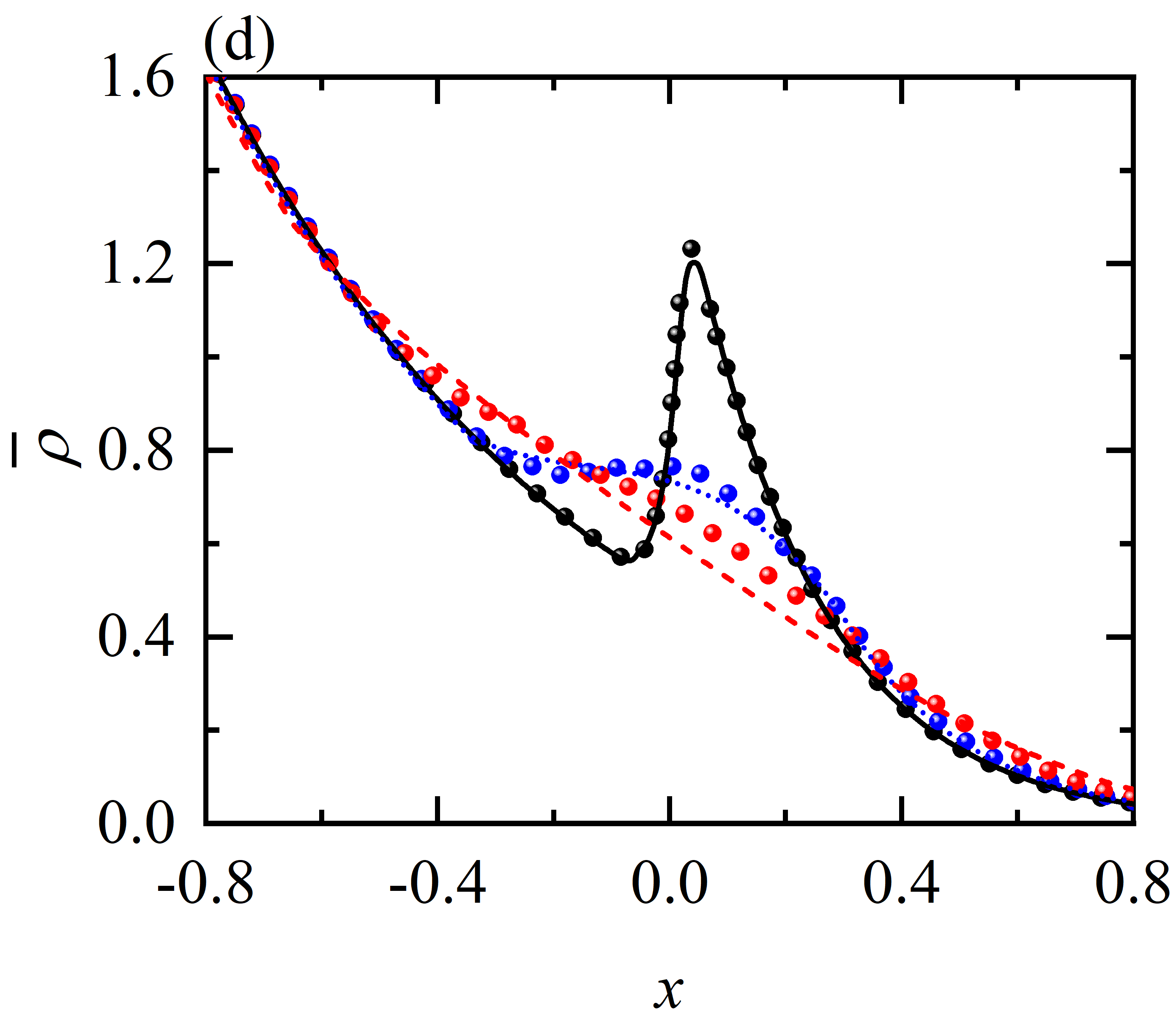}}
\caption{Spatial profiles of the mean density  at three different moments $t/\tau=1$ (colored by black), $t/\tau=2.5$ (colored by blue), and $t/\tau=4$ (colored by red). Subfigures (a) to (d) correspond to the cases with $S_r=$ 0.5, 1, 2, and 3, respectively.  } 
\label{densitypro}
\end{figure}

Except the temporal evolution of mixing width,  spatial profiles of the important variables are also examined. 
Figure \ref{speciespro} presents the mass fraction profiles at three different moments $t/\tau=1$, $t/\tau=2.5$, and $t/\tau=4$, across stratification parameters  $S_r$ varying from 0.5 to 3.
Excellent agreements with HiFi simulations are observed for all cases, demonstrating the model's robust predictive capability.
This validation is further reinforced by the density profile comparisons, as shown in figure \ref{densitypro}, which serve as a critical metric for compressible flow characterization.

The density profile variations  depicted in figure \ref{densitypro} show  critical physical insights, offering a more direct understanding of the baseline model's distinct performance across cases with different $S_r$. 
Specifically, for the cases with a relatively small $S_r$, i.e. $S_r=$0.5 and 1, the sustained positive density gradient throughout the simulation ensures an effective baroclinic production term.
It leads to the baseline model's better performance for mixing width in cases with weaker stratification ($Sr=$0.5 and 1) compared to stronger stratification cases ($Sr=$2 and 3), as shown in figure \ref{w}. 
Additionally, it is also noted that compared to the case of $S_r$=1, the baseline model shows superior accuracy for the case of $S_r$=0.5, as the positive density gradient persists across the entire mixing region, providing a relatively sufficient TKE production term.
However, for the case of $S_r$=1, the positive density gradients become confined to the core mixing region at later stages, while negative gradients emerge at the edge of the mixing region.
It is well-established that the amplitude of the mixing width depends on the penetration dynamics of bubbles and spikes at the edge of the mixing zone.
In RANS simulation,  the strength of penetration is closely linked to the amplitude of the TKE production term.
Due to the disappearance of the baroclinic term at the edge (where the density gradient becomes negative), the TKE production term diminishes to zero, significantly reducing the TKE and halting further development of the mixing evolution.
Consequently, the mixing width predicted by the baseline model is substantially lower than the actual physical evolution.

The underestimate of the mixing width for  cases of $S_r=$2 and 3 can also be  explained by the aforementioned analysis.
For these cases, which exhibit strong density stratification, fundamentally different behavior is observed. 
The density gradient prematurely reverses sign during the simulation---occurring at the edge of the mixing region in the intermediate stage for $S_r$=2 and across the entire mixing region in the intermediate stage for  $S_r$=3.
This gradient reversal causes a severe deficit in the turbulence production source.
Since the baseline model's closure relies solely on the baroclinic term, it fails to provide sufficient TKE production, leading to a deterioration in model performance with increasing stratification strength.

Section \ref{Convective instability in density-stratified compressible RT flow} establishes that flow regions where baroclinic terms  vanish  may maintain instability, as determined by the local instability  criterion for compressible flows.
This phenomenon  has also been validated by HiFi simulations  \citep{Gauthier2017Compressible,Luo2022Mixing}, which demonstrate ongoing mixing and turbulence development in density-stratified compressible RT flows despite baroclinic term disappearance. 
The proposed model incorporates more comprehensive physical mechanisms for compressible mixing flows and   supplements the TKE production source appropriately, leading to successful predictions across all investigated cases.

\section{Summary and discussion}  \label{Summary and discussion} 
Compressible mixing flow induced by hydromechanics interfacial instability play a crucial role in engineering applications, such as the ICF.
RANS simulations remain the most practical approach for engineering predictions.
However, most of the existing RANS mixing models focus on the incompressible cases, and fail to predict compressible mixing flows accurately. 
This study examines the density-stratified RT flows as representative cases to: (i) identify limitations in the existing models for compressible flow predictions, and (ii) develop an improved compressible RANS mixing model. 
Our analysis begins with a comparative study of local instability criteria for incompressible versus compressible flows, demonstrating the physical inadequacy of conventional baroclinic term modeling ($\frac{\partial \bar{p}}{\partial x_{i}}\frac{\partial \bar{\rho}}{\partial x_{i}}$) for compressible cases. 
Subsequently, based on the $K-L-\gamma$ mixing transition model proposed by \citet{Xie2025intermittency}, we derive compressibility corrections for modeling the key turbulent mass flux term in the TKE equation. 
Validation studies across various density-stratified compressible RT cases confirm the model's improved predictive capability.
Compared with the incompressible baseline model, two fundamental improvements account for the proposed model's success.
One is incorporation of the local instability criterion for compressible flows, and the other is proper consideration of turbulent heat flux effect, which prove significant in  flows considered here.
Moreover, the proposed modifications maintain simplicity without introducing additional model coefficient, while substantially enhancing its accuracy for compressible mixing flows.

While elements of the proposed modeling formulation have appeared in previous studies, their application to RANS modeling of interfacial mixing flows remains limited.
The present closure shares some similar modeling with formulations dating back nearly four decades in stellar evolution studies \citep{Cloutman1987New}. 
Furthermore, the kinetic closure model for turbulent counter-gradient diffusion developed by \citet{osti2003Compressibility} provides theoretical support for modeling the turbulent heat flux in this study.
In the existing studies, several researchers have employed partial versions of the current compressibility corrections.
For instance, the $K-\epsilon$ model by \citet{MornLpez2013MulticomponentRN} use a similar form  $\frac{\partial \bar{\rho}}{\partial x_{i}}-\frac{\bar{\rho}}{\bar{p}}\frac{\partial \bar{p}}{\partial x_{i}}$ to close the turbulent mass flux velocity $\overline{u^{''}_{i}}$.
 \citet{Zhou2017ii} have similarly emphasized the importance of entropy gradients in modeling the TKE equation's buoyancy source term.
 However, these pioneering contributions have received relatively little attention in modeling interfacial mixing flows. 
 This may be attributed to the predominantly incompressible or weakly compressible nature of previously studied flows, where compressibility effects are insufficiently pronounced to demonstrate the models' full potential.

Contributions of this study include the following points.
Firstly, we present a systematic derivation of comprehensive compressibility closures based on the  EOS of perfect gas and the thermodynamics law.
This represents a significant advance over previous sporadic and incomplete formulations found in the literature.
The developed closures maintain both physical fidelity and computational tractability.
Second, our results demonstrate the crucial role of turbulent heat flux in compressible flows, while it is ignored in incompressible cases.
Through detailed analysis, it is revealed that turbulent heat flux: (i) significantly modifies buoyancy effects by mediating density fluctuations, and (ii) exhibits counter-gradient diffusion behavior in density-stratified compressible RT flows. 
The latter has not been uncovered in the existing reports, to our knowledge.
It necessitate specific modifications, while the traditional GDA fails to describe that.
Thirdly, this work provides the first successful RANS-based framework for predicting density-stratified compressible RT flows, representing a significant step forward in turbulence modeling for practical engineering applications.

While the proposed model shows promising results in the density-stratified compressible RT mixing flows, its performance for the compressible RM and KH mixing problems remains to be   verified in the future.
The present study primarily establishes a foundational modeling strategy for compressible mixing flows, with comprehensive verification left for future investigation.

\backsection[Acknowledgements]
 
This work was supported by the National Natural Science Foundation of China (Grant Nos. 12222203, 92152202, and 12432010) and the National Key Laboratory of Computational Physics (Grant No. 6142A05240201). Additionally, Luo, T.  was supported by  the Research Grants Council (RGC) of the Hong Kong Special Administrative Region, China (Project Reference: AoE/P‐601/23‐N) and the Center for Ocean Research in Hong Kong and Macau (CORE), a joint research center between the Laoshan Laboratory and the Hong Kong University of Science and Technology (HKUST). 

\backsection[Declaration of Interests]
 
The authors report no conflict of interest.

\bibliographystyle{jfm}
\bibliography{jfm}

\begin{thebibliography}{48}
\expandafter\ifx\csname natexlab\endcsname\relax\def\natexlab#1{#1}\fi
\def\au#1{#1} \def\ed#1{#1} \def\yr#1{#1}\def\at#1{#1}\def\jt#1{\textit{#1}} \def\bt#1{#1}\def\bvol#1{\textbf{#1}} \def\vol#1{#1} \def\pg#1{#1} \def\publ#1{#1}\def\arxiv#1{#1}\def\org#1{#1}\def\st#1{\textit{#1}}

\bibitem[Chandrasekhar(1961)]{Chandrasekhar1961Hydrodynamic}
{\sc \au{Chandrasekhar, S.}} \yr{1961} {\em Hydrodynamic and hydromagnetic stability\/}.  \publ{Oxford-{C}larendon {P}ress and {N}ew {Y}ork-{O}xford {U}niv. {P}ress}.

\bibitem[Chassaing {\em et~al.\/}(2002)Chassaing, Antonia, Anselmet, Joly \& Sarkar]{Chassaing2002Variable}
{\sc \au{Chassaing, P.}, \au{Antonia, R.}, \au{Anselmet, F.}, \au{Joly, L.} \& \au{Sarkar, S.}} \yr{2002} {\em Variable Density Fluid Turbulence\/}.  \publ{Springer Netherlands: Imprint: Springer}.

\bibitem[Cloutman(1987)]{Cloutman1987New}
{\sc \au{Cloutman, L.~D.}} \yr{1987}  \at{A new estimate of the mixing length and convective overshooting in massive stars}.  \jt{Astrophys. J.}  \bvol{313},  \pg{699--710}.

\bibitem[Cloutman(2003)]{osti2003Compressibility}
{\sc \au{Cloutman, L.~D.}} \yr{2003}  \bt{Compressibility corrections to closure approximations for turbulent flow simulations}. {\em Tech. Rep.\/}.  \org{Lawrence Livermore National Lab. (LLNL), Livermore, CA (United States)}.

\bibitem[Craxton {\em et~al.\/}(2015)Craxton, Anderson, Boehly, Goncharov, Harding, Knauer, McCrory, McKenty, Meyerhofer, Myatt, Schmitt, Sethian, Short, Skupsky, Theobald, Kruer, Tanaka, Betti, Collins, Delettrez, Hu, Marozas, Maximov, Michel, Radha, Regan, Sangster, Seka, Solodov, Soures, Stoeckl \& Zuegel]{Craxton2015Direct}
{\sc \au{Craxton, R.~S.}, \au{Anderson, K.~S.}, \au{Boehly, T.~R.}, \au{Goncharov, V.~N.}, \au{Harding, D.~R.}, \au{Knauer, J.~P.}, \au{McCrory, R.~L.}, \au{McKenty, P.~W.}, \au{Meyerhofer, D.~D.}, \au{Myatt, J.~F.}, \au{Schmitt, A.~J.}, \au{Sethian, J.~D.}, \au{Short, R.~W.}, \au{Skupsky, S.}, \au{Theobald, W.}, \au{Kruer, W.~L.}, \au{Tanaka, K.}, \au{Betti, R.}, \au{Collins, T. J.~B.}, \au{Delettrez, J.~A.}, \au{Hu, S.~X.}, \au{Marozas, J.~A.}, \au{Maximov, A.~V.}, \au{Michel, D.~T.}, \au{Radha, P.~B.}, \au{Regan, S.~P.}, \au{Sangster, T.~C.}, \au{Seka, W.}, \au{Solodov, A.~A.}, \au{Soures, J.~M.}, \au{Stoeckl, C.} \& \au{Zuegel, J.~D.}} \yr{2015}  \at{Direct-drive inertial confinement fusion: {A} review}.  \jt{Phys. Plasmas}  \bvol{22}~(11),  \pg{110501}.

\bibitem[Denissen {\em et~al.\/}(2014)Denissen, Rollin, Reisner \& Andrews]{Denissen2014tilted}
{\sc \au{Denissen, N.}, \au{Rollin, B.}, \au{Reisner, J.} \& \au{Andrews, M.}} \yr{2014}  \at{The tilted rocket rig: A {R}ayleigh-{T}aylor test case for {RANS} models}.  \jt{J. Fluids Eng.}  \bvol{136},  \pg{091301}.

\bibitem[Dimonte \& Tipton(2006)]{Dimonte2006K-L}
{\sc \au{Dimonte, G.} \& \au{Tipton, R.}} \yr{2006}  \at{{K-L} turbulence model for the self-similar growth of the {R}ayleigh-{T}aylor and {R}ichtmyer-{M}eshkov instabilities}.  \jt{Phys. Fluids}  \bvol{18}~(8),  \pg{85101}.

\bibitem[Fu {\em et~al.\/}(2022)Fu, Zhao, Xu, Wang, Liu, Wan \& Lu]{Fu2022Nonlinear}
{\sc \au{Fu, Chengquan}, \au{Zhao, Zhiye}, \au{Xu, Xin}, \au{Wang, Pei}, \au{Liu, Nansheng}, \au{Wan, Zhenhua} \& \au{Lu, Xiyun}} \yr{2022}  \at{Nonlinear saturation of bubble evolution in a two-dimensional single-mode stratified compressible {R}ayleigh-{T}aylor instability}.  \jt{Phys. Rev. Fluids}  \bvol{7},  \pg{023902}.

\bibitem[Gamalii {\em et~al.\/}(1980)Gamalii, Rozanov, Samarskii, Tishkin, Tyurina \& Favorskii]{Gamalii1980Hydrodynamic}
{\sc \au{Gamalii, E.~G.}, \au{Rozanov, V.~B.}, \au{Samarskii, A.~A.}, \au{Tishkin, V.~F.}, \au{Tyurina, N.~N.} \& \au{Favorskii, A.~P.}} \yr{1980}  \at{Hydrodynamic stability of compression of spherical laser targets}.  \jt{Sov. Phys. {JETP}}  \bvol{52},  \pg{230--237}.

\bibitem[Gauthier(2013)]{Gauthier2013Compressibility}
{\sc \au{Gauthier, S.}} \yr{2013}  \at{Compressibility effects in {R}ayleigh-{T}aylor flows: influence of the stratification}.  \jt{Phys. Scr.}  \bvol{2013}~(T155),  \pg{014012}.

\bibitem[Gauthier(2017)]{Gauthier2017Compressible}
{\sc \au{Gauthier, Serge}} \yr{2017}  \at{Compressible {R}ayleigh-{T}aylor turbulent mixing layer between newtonian miscible fluids}.  \jt{J. Fluid Mech.}  \bvol{830},  \pg{211–256}.

\bibitem[Gauthier \& Le~Creurer(2010)]{Gauthier2010Compressibility}
{\sc \au{Gauthier, S.} \& \au{Le~Creurer, B.}} \yr{2010}  \at{Compressibility effects in {R}ayleigh-{T}aylor instability-induced flows}.  \jt{Philos. T. R. Soc. A}  \bvol{368}~(1916),  \pg{1681--1704}.

\bibitem[G\'{e}nin \& Menon(2010)]{Franklin2010Simulation}
{\sc \au{G\'{e}nin, F.} \& \au{Menon, S.}} \yr{2010}  \at{Simulation of turbulent mixing behind a strut injector in supersonic flow}.  \jt{AIAA J.}  \bvol{48}~(3),  \pg{526--539}.

\bibitem[George \& Glimm(2005)]{George2005Self}
{\sc \au{George, E.} \& \au{Glimm, J.}} \yr{2005}  \at{Self-similarity of {R}ayleigh-{T}aylor mixing rates}.  \jt{Phys. Fluids}  \bvol{17}~(5),  \pg{054101}.

\bibitem[Helmholtz(1868)]{helmholtz1868on}
{\sc \au{Helmholtz, V.}} \yr{1868}  \at{On discontinuous movements of fluids}.  \jt{Lond. Edinb. Dubl. Phil. Mag. J. Sci.}  \bvol{36}~(244),  \pg{337--346}.

\bibitem[Jin {\em et~al.\/}(2005)Jin, Liu, Lu, Cheng, Glimm \& Sharp]{Jin2005Rayleigh}
{\sc \au{Jin, H.}, \au{Liu, X.~F.}, \au{Lu, T.}, \au{Cheng, B.}, \au{Glimm, J.} \& \au{Sharp, D.~H.}} \yr{2005}  \at{Rayleigh-{T}aylor mixing rates for compressible flow}.  \jt{Phys. Fluids}  \bvol{17}~(2),  \pg{024104}.

\bibitem[Kelvin(1871)]{kelvin1871hydrokinetic}
{\sc \au{Kelvin, L.}} \yr{1871}  \at{Hydrokinetic solutions and observations}.  \jt{Lond. Edinb. Dubl. Phil. Mag. J. Sci.}  \bvol{42}~(281),  \pg{362--377}.

\bibitem[Kokkinakis {\em et~al.\/}(2019)Kokkinakis, Drikakis \& Youngs]{Kokkinakis2019Modeling}
{\sc \au{Kokkinakis, I.~W.}, \au{Drikakis, D.} \& \au{Youngs, D.~L.}} \yr{2019}  \at{Modeling of {R}ayleigh-{T}aylor mixing using single-fluid models}.  \jt{Phys. Rev. E}  \bvol{99},  \pg{013104}.

\bibitem[Kokkinakis {\em et~al.\/}(2020)Kokkinakis, Drikakis \& Youngs]{Kokkinakis2020Two}
{\sc \au{Kokkinakis, I.~W.}, \au{Drikakis, D.} \& \au{Youngs, D.~L.}} \yr{2020}  \at{Two-equation and multi-fluid turbulence models for {R}ichtmyer-{M}eshkov mixing}.  \jt{Phys. Fluids}  \bvol{32}~(7),  \pg{074102}.

\bibitem[Kokkinakis {\em et~al.\/}(2015)Kokkinakis, Drikakis, Youngs \& Williams]{Kokkinakis2015Two}
{\sc \au{Kokkinakis, I.~W.}, \au{Drikakis, D.}, \au{Youngs, D.~L.} \& \au{Williams, R. J.~R.}} \yr{2015}  \at{Two-equation and multi fluid turbulence models for {R}ayleigh-{T}aylor mixing}.  \jt{Int. J. Heat Fluid Fl.}  \bvol{56},  \pg{233--250}.

\bibitem[Li {\em et~al.\/}(2019)Li, He, Zhang \& Tian]{Li2019On}
{\sc \au{Li, Haifeng}, \au{He, Zhiwei}, \au{Zhang, Yousheng} \& \au{Tian, Baolin}} \yr{2019}  \at{On the role of rarefaction/compression waves in {R}ichtmyer-{M}eshkov instability with reshock}.  \jt{Phys. Fluids}  \bvol{31}~(5),  \pg{054102}.

\bibitem[Livescu(2004)]{Livescu2004Compressibility}
{\sc \au{Livescu, D.}} \yr{2004}  \at{Compressibility effects on the {R}ayleigh-{T}aylor instability growth between immiscible fluids}.  \jt{Phys. Fluids}  \bvol{16}~(1),  \pg{118--127}.

\bibitem[Livescu(2013)]{Livescu2013Num}
{\sc \au{Livescu, D.}} \yr{2013}  \at{Numerical simulations of two-fluid turbulent mixing at large density ratios and applications to the {R}ayleigh-{T}aylor instability}.  \jt{Philos. Trans. R. Soc. A}  \bvol{371}~(2003),  \pg{20120185}.

\bibitem[Luo {\em et~al.\/}(2024)Luo, Li, Yuan, Peng, Liu, Wang \& Wang]{Luo2024Fourier}
{\sc \au{Luo, Tengfei}, \au{Li, Zhijie}, \au{Yuan, Zelong}, \au{Peng, Wenhui}, \au{Liu, Tianyuan}, \au{Wang, Liangzhu~(Leon)} \& \au{Wang, Jianchun}} \yr{2024}  \at{Fourier neural operator for large eddy simulation of compressible {R}ayleigh-{T}aylor turbulence}.  \jt{Phys. Fluids}  \bvol{36}~(7),  \pg{075165}.

\bibitem[Luo \& Wang(2022)]{Luo2022Mixing}
{\sc \au{Luo, Tengfei} \& \au{Wang, Jianchun}} \yr{2022}  \at{Mixing and energy transfer in compressible {R}ayleigh-{T}aylor turbulence for initial isothermal stratification}.  \jt{Phys. Rev. Fluids}  \bvol{7},  \pg{104608}.

\bibitem[Meshkov(1969)]{meshkov1969instability}
{\sc \au{Meshkov, E.~E.}} \yr{1969}  \at{Instability of the interface of two gases accelerated by a shock wave}.  \jt{Fluid Dyn.}  \bvol{4}~(5),  \pg{101--104}.

\bibitem[Mor{\'a}n-L{\'o}pez \& Schilling(2013)]{MornLpez2013MulticomponentRN}
{\sc \au{Mor{\'a}n-L{\'o}pez, J.} \& \au{Schilling, O.}} \yr{2013}  \at{Multicomponent {R}eynolds-averaged {N}avier-{S}tokes simulations of reshocked {R}ichtmyer-{M}eshkov instability-induced mixing}.  \jt{High Energy Density Phys.}  \bvol{9},  \pg{112--121}.

\bibitem[Morgan \& Greenough(2015)]{Morgan2015Large}
{\sc \au{Morgan, B.} \& \au{Greenough, J.}} \yr{2015}  \at{Large-eddy and unsteady {RANS} simulations of a shock-accelerated heavy gas cylinder}.  \jt{Shock Waves}  \bvol{26}.

\bibitem[Qi {\em et~al.\/}(2024)Qi, He, Xu \& Zhang]{Qi2024vortex}
{\sc \au{Qi, Han}, \au{He, Zhi-wei}, \au{Xu, Ai-guo} \& \au{Zhang, You-sheng}} \yr{2024}  \at{The vortex structure and enstrophy of the mixing transition induced by {R}ayleigh-{T}aylor instability}.  \jt{Phys. Fluids}  \bvol{36}~(11),  \pg{114107}.

\bibitem[Rayleigh(1882)]{rayleigh1882investigation}
{\sc \au{Rayleigh, L.}} \yr{1882}  \at{Investigation of the {C}haracter of the {E}quilibrium of an {I}ncompressible {H}eavy {F}luid of {V}ariable {D}ensity}.  \jt{Proc. Lond. Math. Soc.}  \bvol{201}~(1),  \pg{170--177}.

\bibitem[Richtmyer(1960)]{richtmyer1960taylor}
{\sc \au{Richtmyer, R.~D.}} \yr{1960}  \at{Taylor instability in shock acceleration of compressible fluids}.  \jt{Commun. Pure Appl. Math.}  \bvol{13}~(2),  \pg{297--319}.

\bibitem[Taylor(1950)]{taylor1950instability}
{\sc \au{Taylor, G.}} \yr{1950}  \at{The {I}nstability of {L}iquid {S}urfaces when {A}ccelerated in a {D}irection {P}erpendicular to their {P}lanes. {I}}.  \jt{Proc. R. Soc. Lond., Ser. A}  \bvol{201}~(1065),  \pg{192}.

\bibitem[Thornber {\em et~al.\/}(2008{\natexlab{{\em a\/}}})Thornber, Drikakis, Williams \& Youngs]{Thornber2008on}
{\sc \au{Thornber, B.}, \au{Drikakis, D.}, \au{Williams, R. J.~R.} \& \au{Youngs, D.}} \yr{2008{\natexlab{{\em a\/}}}}  \at{On entropy generation and dissipation of kinetic energy in high-resolution shock-capturing schemes}.  \jt{J. Comput. Phys.}  \bvol{227}~(10),  \pg{4853--4872}.

\bibitem[Thornber {\em et~al.\/}(2008{\natexlab{{\em b\/}}})Thornber, Mosedale, Drikakis, Youngs \& Williams]{Thornber2008An}
{\sc \au{Thornber, B.}, \au{Mosedale, A.}, \au{Drikakis, D.}, \au{Youngs, D.} \& \au{Williams, R. J.~R.}} \yr{2008{\natexlab{{\em b\/}}}}  \at{An improved reconstruction method for compressible flows with low mach number features}.  \jt{J. Comput. Phys.}  \bvol{227}~(10),  \pg{4873--4894}.

\bibitem[Toro {\em et~al.\/}(1994)Toro, Spruce \& Speares]{Toro1994RestorationOT}
{\sc \au{Toro, E.~F.}, \au{Spruce, M.} \& \au{Speares, W.}} \yr{1994}  \at{Restoration of the contact surface in the {HLL}-{R}iemann solver}.  \jt{Shock Waves}  \bvol{4},  \pg{25--34}.

\bibitem[Xiao {\em et~al.\/}(2022)Xiao, Hu, Dai \& Zhang]{Xiao2022exp}
{\sc \au{Xiao, M.}, \au{Hu, Z.}, \au{Dai, Z.} \& \au{Zhang, Y.}} \yr{2022}  \at{Experimentally consistent large-eddy simulation of re-shocked {R}ichtmyer-{M}eshkov turbulent mixing}.  \jt{Phys. Fluids}  \bvol{34}~(12), 125125.

\bibitem[Xiao {\em et~al.\/}(2025)Xiao, Qi \& Zhang]{Xiao2025Local}
{\sc \au{Xiao, Mengjuan}, \au{Qi, Han} \& \au{Zhang, Yousheng}} \yr{2025}  \at{Local transition indicator and modelling of turbulent mixing based on the mixing state}.  \jt{J. Fluid Mech.}  \bvol{1002},  \pg{A4}.

\bibitem[Xiao {\em et~al.\/}(2020)Xiao, Zhang \& Tian]{Xiao2020Modeling}
{\sc \au{Xiao, M.}, \au{Zhang, Y.} \& \au{Tian, B.}} \yr{2020}  \at{Modeling of turbulent mixing with an improved {K-L} model}.  \jt{Phys. Fluids}  \bvol{32}~(9),  \pg{092104}.

\bibitem[Xie {\em et~al.\/}(2025)Xie, Qi, Xiao, Zhang \& Zhao]{Xie2025intermittency}
{\sc \au{Xie, Hansong}, \au{Qi, Han}, \au{Xiao, Mengjuan}, \au{Zhang, Yousheng} \& \au{Zhao, Yaomin}} \yr{2025}  \at{An intermittency based {R}eynolds-averaged transition model for mixing flows induced by interfacial instabilities}.  \jt{J. Fluid Mech.}  \bvol{1002},  \pg{A31}.

\bibitem[Xie {\em et~al.\/}(2023)Xie, Zhao \& Zhang]{Xie2023Data}
{\sc \au{Xie, H.}, \au{Zhao, Y.} \& \au{Zhang, Y.}} \yr{2023}  \at{Data-driven nonlinear {K-L} turbulent mixing model via gene expression programming method}.  \jt{Acta Mech. Sin.}  \bvol{39}~(2),  \pg{322315}.

\bibitem[Xie {\em et~al.\/}(2021{\natexlab{{\em a\/}}})Xie, Xiao \& Zhang]{Xie2021Predicting}
{\sc \au{Xie, Han-song}, \au{Xiao, Meng-juan} \& \au{Zhang, You-sheng}} \yr{2021{\natexlab{{\em a\/}}}}  \at{Predicting different turbulent mixing problems with the same k-$\epsilon$ model and model coefficients}.  \jt{AIP Advances}  \bvol{11}~(7),  \pg{075213}.

\bibitem[Xie {\em et~al.\/}(2021{\natexlab{{\em b\/}}})Xie, Xiao \& Zhang]{Xie2021Unified}
{\sc \au{Xie, H.-S.}, \au{Xiao, M.-J.} \& \au{Zhang, Y.-S.}} \yr{2021{\natexlab{{\em b\/}}}}  \at{Unified prediction of turbulent mixing induced by interfacial instabilities via {B}esnard-{H}arlow-{R}auenzahn-2 model}.  \jt{Phys. Fluids}  \bvol{33}~(10),  \pg{105123}.

\bibitem[Xue \& Ye(2010)]{Xue2010Destabilizing}
{\sc \au{Xue, Chuang} \& \au{Ye, Wenhua}} \yr{2010}  \at{Destabilizing effect of compressibility on {R}ayleigh-{T}aylor instability for fluids with fixed density profile}.  \jt{Phys. Plasmas}  \bvol{17}~(4),  \pg{042705}.

\bibitem[Zhang {\em et~al.\/}(2020)Zhang, He, Xie, Xiao \& Tian]{zhang2020Methodology}
{\sc \au{Zhang, Y.}, \au{He, Z.}, \au{Xie, H.}, \au{Xiao, M.} \& \au{Tian, B.}} \yr{2020}  \at{Methodology for determining coefficients of turbulent mixing model}.  \jt{J. Fluid Mech.}  \bvol{905},  \pg{A26}.

\bibitem[Zhao {\em et~al.\/}(2022)Zhao, Betti \& Aluie]{Zhao2022Scale}
{\sc \au{Zhao, D.}, \au{Betti, R.} \& \au{Aluie, H.}} \yr{2022}  \at{Scale interactions and anisotropy in {R}ayleigh-{T}aylor turbulence}.  \jt{J. Fluid Mech.}  \bvol{930},  \pg{A29}.

\bibitem[Zhao {\em et~al.\/}(2020)Zhao, Liu \& Lu]{Zhao2020Kinetic}
{\sc \au{Zhao, Zhiye}, \au{Liu, Nansheng} \& \au{Lu, Xiyun}} \yr{2020}  \at{Kinetic energy and enstrophy transfer in compressible {R}ayleigh-{T}aylor turbulence}.  \jt{J. Fluid Mech.}  \bvol{904},  \pg{A37}.

\bibitem[Zhou(2017{\natexlab{{\em a\/}}})]{zhou2017Rayleigh}
{\sc \au{Zhou, Y.}} \yr{2017{\natexlab{{\em a\/}}}}  \at{Rayleigh-{T}aylor and {R}ichtmyer-{M}eshkov instability induced flow, turbulence, and mixing. {I}}.  \jt{Phys. Rep.}  \bvol{720-722},  \pg{1--136}.

\bibitem[Zhou(2017{\natexlab{{\em b\/}}})]{Zhou2017ii}
{\sc \au{Zhou, Y.}} \yr{2017{\natexlab{{\em b\/}}}}  \at{Rayleigh-{T}aylor and {R}ichtmyer-{M}eshkov instability induced flow, turbulence, and mixing. {II}}.  \jt{Phys. Rep.}  \bvol{723-725},  \pg{1--160}.

\end{thebibliography}

\end{document}